\documentclass[noshowpacs,pra,aps,superscriptaddress,floatfix,twocolumn]{revtex4}

\usepackage{lineno,hyperref,mathtools,doi}
\usepackage{amsmath}
\usepackage{amssymb,mathrsfs}
\usepackage{graphicx}
\usepackage{hyperref}
\usepackage[dvipsnames]{xcolor}

\newcommand{\prt}{\partial}
\newcommand{\la}{\lambda}
\newcommand{\sss}{\scriptscriptstyle}
\newcommand{\xsol}{x_{\sss\rm S}}
\newcommand{\lsol}{\lambda_{\sss\rm S}}

\begin{document}
\title{Wave breaking and formation of dispersive shock waves in a
  defocusing nonlinear optical material}

\author{M. Isoard} \affiliation{LPTMS, UMR 8626, CNRS,
  Univ. Paris-Sud, Universit\'e Paris-Saclay, 91405 Orsay, France}
\author{A. M. Kamchatnov} \affiliation{Institute of Spectroscopy,
  Russian Academy of Sciences, Troitsk, Moscow, 108840, Russia}
\affiliation{Moscow Institute of Physics and Technology, Institutsky
  lane 9, Dolgoprudny, Moscow region, 141701, Russia}
\author{N. Pavloff} \affiliation{LPTMS, UMR 8626, CNRS,
  Univ. Paris-Sud, Universit\'e Paris-Saclay, 91405 Orsay, France}

\begin{abstract}
  We theoretically describe the quasi one-dimensional transverse
  spreading of a light beam propagating in a nonlinear optical
  material in the presence of a uniform background light
  intensity. For short propagation distances the pulse can be
  described within a nondispersive (geometric optics) approximation by
  means of Riemann's approach. For larger distances, wave breaking
  occurs, leading to the formation of dispersive shocks at both edges
  of the beam.  We describe this phenomenon within Whitham modulation
  theory, which yields an excellent agreement with numerical
  simulations. Our analytic approach makes it possible to extract the
  leading asymptotic behavior of the parameters of the shock, setting
  up the basis for a theory of non-dissipative weak shocks.
\end{abstract}

\maketitle

\section{Introduction}\label{intro}

It has long been realized that light propagating in a nonlinear medium
was amenable to a hydrodynamic treatment, see e.g.,
Refs.~\cite{Tal65,ASK66,ASK67}. In the case of a defocusing
nonlinearity, this rich analogy has not only triggered experimental
research, but also made it possible to get an intuitive understanding
of observations such as the formation of rings in the far field beyond
a nonlinear slab \cite{Akh67,Har97}, of dark solitons
\cite{Emp87,Kro88,Wei88}, vortices \cite{Are91,Vau96,Voc18}, wave
breaking \cite{Rot89,Gho07} and dispersive shock waves
\cite{Cou04,wkf-07,Bar07,Fat14,Xu16,Xu17}, of spontaneously
self-accelerated Airy beams \cite{Kar15}, of optical event horizon
\cite{Ela12}, ergo-regions \cite{Voc17} and stimulated Hawking
radiation \cite{Dro18}, of sonic-like dispersion relation
\cite{Voc15,Fon18} and superfluid motion \cite{Mic18}.  Very similar
phenomena have also been observed in the neighboring fields of cavity
polaritons and of Bose-Einstein condensation of atomic vapors. They
all result from the interplay between nonlinearity and dispersion,
whose effects become prominent near a gradient catastrophe region.

In this work we present a theoretical treatment of a model
configuration which has been realized experimentally in a
one-dimensional situation in Refs.~\cite{wkf-07,Xu16}: the nonlinear
spreading of a region of increased light intensity in the presence of
a uniform constant background. In the absence of background, and for a
smooth initial intensity pattern, the spreading is mainly driven by
the nonlinear defocusing and can be treated analytically in some
simple cases \cite{Tal65}. The situation is more interesting in the
presence of a constant background: the pulse splits in two parts, each
eventually experiencing nonlinear wave breaking, leading to the
formation of a dispersive shock wave (DSW) which cannot be described
within the framework of perturbation theory, even if the region of
increased intensity corresponds to a weak perturbation of the flat
pedestal. This scenario indeed fits with the hydrodynamic approach of
nonlinear light propagation, and is nicely confirmed by the
experimental observations of Refs.~\cite{wkf-07,Xu16}. Although the
numerical treatment of the problem is relatively simple
\cite{kgk-04,El07,Conf12}, a theoretical approach to both the initial
splitting of the pulse and the subsequent shock formation requires a
careful analysis. The goal of this article is to present such an
analysis. A most significant outcome of our detailed treatment is a
simple asymptotic description of some important shock parameters.
This provides a non-dissipative counterpart of the usual weak viscous
shock theory (see, e.g., Ref.~\cite{whitham-74}) and paves the way for
a quantitative experimental test of our predictions.

The paper is organized as follows: In Sec.~\ref{sec:model} we present
the model and the set-up we aim at studying. After a brief discussion
of shortcomings of the linearized approach, the spreading and
splitting stage of evolution is accounted for in Sec.~\ref{DSE} within
a dispersionless approximation which holds when the pulse region
initially presents no large intensity gradient. It is well known that
in such a situation the light flow can be described by
hydrodynamic-like equations which can be cast into a diagonal form for
two new position and time-dependent variables --- the so called
Riemann invariants. The difficulty here lies in the fact that the
splitting involves simultaneous variations of both of them: one does
not have an initial simple wave within which one of the Riemann
invariants remains constant, as occurs for instance in a similar
uni-directional propagation case modeled by the Korteweg-de Vries
equation (see, e.g., Ref.~\cite{Iso18}). We treat the problem in
Secs.~\ref{sec.EP} and \ref{sec.match} using an extension of the
Riemann method due to Ludford \cite{Lud52} (also used in
Ref.~\cite{For2009}) and compare the results with numerical
simulations in Sec.~\ref{sec.num}. During the spreading of the pulse,
nonlinear effects induce wave steepening which results in a gradient
catastrophe and wave breaking. After the wave breaking time,
dispersive effects can no longer be omitted, a shock is formed, and in
this case we resort to Whitham modulation theory \cite{whitham-74} to
describe the time evolution of the pulse. Such a treatment was
initiated long ago by Gurevich and Pitaevskii \cite{gp-73}, and since
that time it has developed into a powerful method with numerous
applications (see, e.g., the review article \cite{eh-16}). Here there
is an additional complexity which lies---as for the initial
non-dispersive stage of evolution---in the fact that two of the (now
four) Riemann invariants which describe the modulated nonlinear
oscillations vary in the shock region. Such a wave has been termed as
``quasi-simple'' in Ref.~\cite{gkm-89}, and a thorough treatment
within Whitham theory has been achieved in the Korteweg-de Vries case
in Refs.~\cite{Gur91,Gur92,Kry92,ek-93}. We generalize in
Sec.~\ref{WGH} this approach to the nonlinear Schrodinger equation
(NLS) which describes light propagation in the nonlinear Kerr medium
(see also Ref.~\cite{El2009}). An interesting outcome of our
theoretical treatment is the asymptotic determination of
experimentally relevant parameters of the dispersive shock, see
Sec.~\ref{sol-edge}. In Sec.~\ref{full} we present the full Whitham
treatment of the after-shock evolution and compare the theoretical
results with numerical simulations. We present in Sec. \ref{discu} a
panorama the different regimes we have identified, and discuss how our
approach can be used to get a simple estimate of the contrast of the
fringes of the DSW. This should be helpful for determining the best
experimental configuration for studying the wave breaking phenomenon
and the subsequent dispersive shock. Our conclusions are presented in
Sec.~\ref{conclusion}.

\section{The model and the linear approximation}\label{sec:model}

In the paraxial approximation, the stationary propagation of the
complex amplitude $A(\vec{r}\,)$ of the electric field of a
monochromatic beam is described by the equation (see, e.g., Ref.~\cite{LL8})
\begin{equation}
{\rm i}\partial_z A = -\frac{1}{2 n_0 k_0} \vec{\nabla}^2_{\!\perp} A
-k_0 \delta n\, A\; .
\end{equation}
In this equation, $n_0$ is the linear refractive index,
$k_0=2\pi/\lambda_0$ is the carrier wave vector, $z$ is the coordinate
along the beam, $\vec{\nabla}^2_\perp$ the transverse Laplacian and
$\delta n$ is a nonlinear contribution to the index. In a non
absorbing defocusing Kerr nonlinear medium one can write
$\delta n=-n_2 |A|^2$, with $n_2>0$.

We define dimensionless units by choosing a reference intensity
$I_{\rm ref}$ and introducing the nonlinear length
$z_{\rm\sss NL}=(k_0 n_2 I_{\rm ref})^{-1}$ and the transverse healing
length $\xi_\perp=(z_{\rm\sss NL}/n_0 k_0)^{1/2}$. We consider a
geometry where the transverse profile is translationally invariant and
depends on a single Cartesian coordinate. One thus writes
$\vec{\nabla}^2_{\!\perp} = \xi_\perp^{-2} \partial_x^2$ where $x$
is the dimensionless transverse coordinate and we define an effective
``time'' $t=z/z_{\rm\sss NL}$. The quantity
$\psi(x,t)=A/\sqrt{I_{\rm ref}}$ is then a solution of the dimensionless
NLS equation
\begin{equation}\label{eq:nls}
{\rm i}\,\psi_t=-\tfrac12 \psi_{xx}+|\psi|^2\psi\; .
\end{equation}
In the following we consider a system with a uniform background light
intensity, on top of which an initial pulse is added at the entrance
of the nonlinear cell. The initial $\psi(x,t=0)$ is real (i.e., no
transverse velocity or, in optical context, no focusing of the light
beam at the input plane), with a dimensionless intensity
$\rho(x,t)=|\psi|^2$ which departs from the constant background value
(which we denote as $\rho_0$) only in the region near the origin where
it forms a bump. To be specific, we consider the typical case where
\begin{equation}\label{rho_init}
\rho(x,0)=\begin{cases}
\rho_0+\rho_1 (1-x^2/x_0^2) & \mbox{if}\quad |x|<x_0, \\
\rho_0 & \mbox{if}\quad |x|\ge x_0.\end{cases}
\end{equation}
We will denote as $\rho_{m}=\rho_0+\rho_1$ the maximal density of
the initial profile. It would be natural to choose the reference light
intensity $I_{\rm ref}$ to be equal to the background one, in this
case one would have $\rho_0=1$. However, we prefer to
be more general and to allow for values of $\rho_0$ different from
unity.

\begin{figure}
\centering
\includegraphics[width=\linewidth]{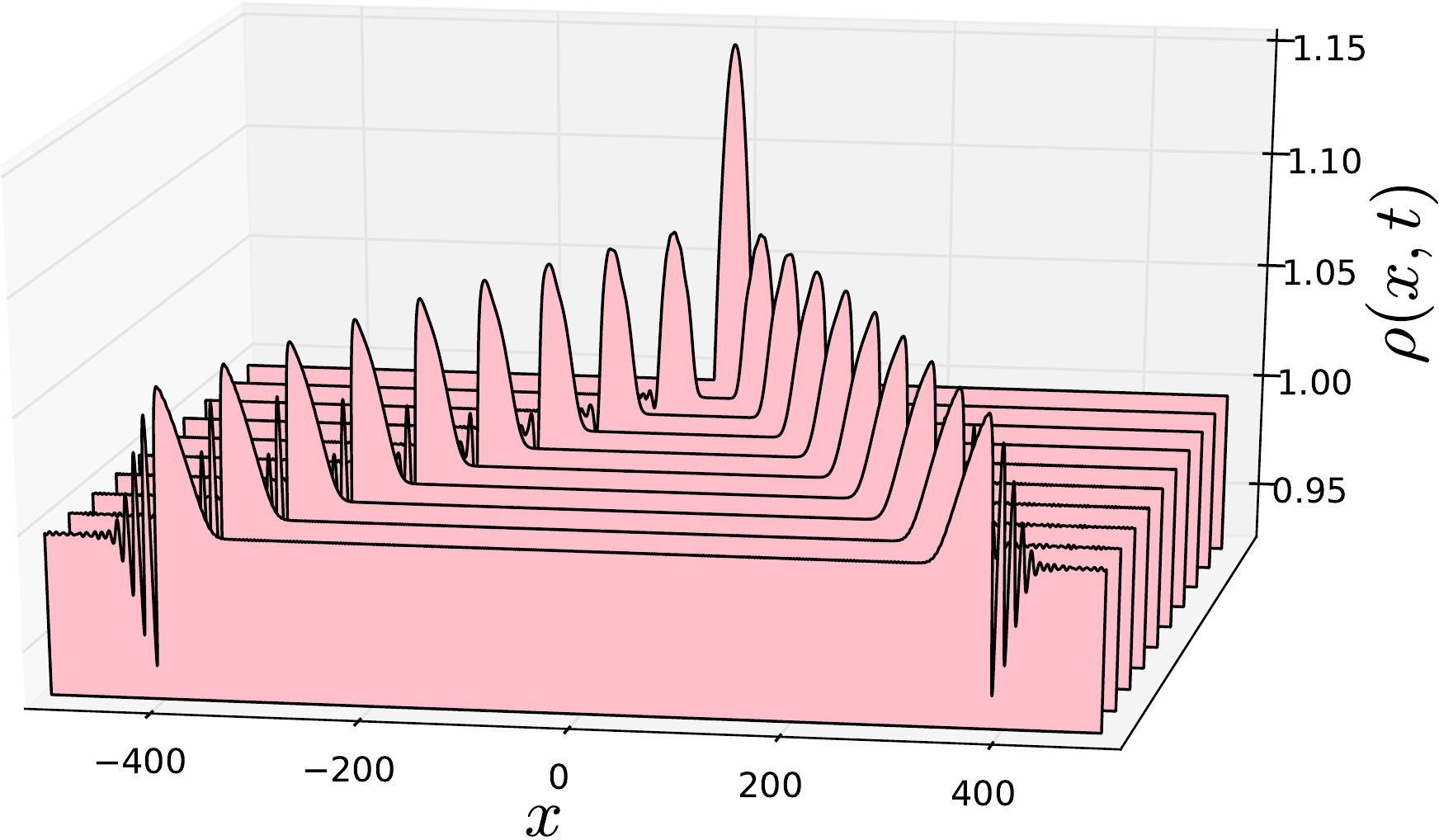}
\includegraphics[width=\linewidth]{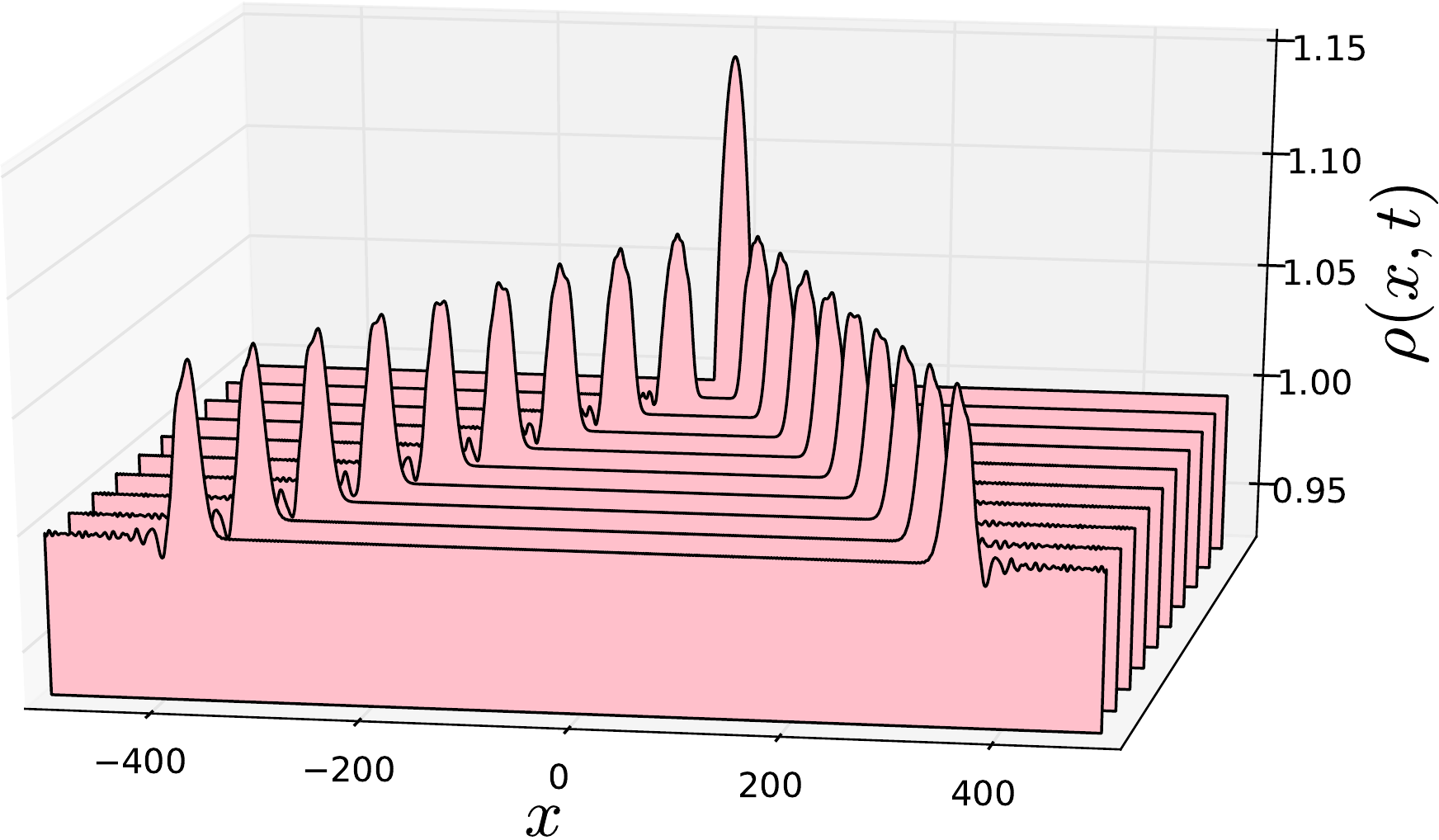}
\caption{Density profiles $\rho(x,t)$ for an initial condition
  $u(x,0)=0$ and $\rho(x,0)$ given by \eqref{rho_init} with
  $\rho_0=1$, $\rho_1=0.15$ and $x_0=20$. The upper part of the figure
  represents the results of the numerical solution of
  Eq.~\eqref{eq:nls}, while the lower part corresponds to the
  linearized version \eqref{eq:linNLS}. The profiles are plotted from time
  $t=0$ to $t=360$ with a time step equal to 40.}
\label{fig1}
\end{figure}
We stress here the paramount importance of nonlinear effects at large
``time'' (i.e., for large propagation distance in the nonlinear
medium). Even for a bump which weakly departs from the background
density, a perturbative approach fails after the wave breaking
time. This is illustrated in Fig.~\ref{fig1} which compare numerical
simulations of the full equation \eqref{eq:nls} with its linearized
version. The linearized treatment is obtained by writing
$\psi(x,t)=\exp(-{\rm i} \rho_0 t)(\sqrt{\rho_0}+\delta\psi(x,t))$ and
assuming that $|\delta\psi|^2\ll \rho_0$ which yields the following
evolution equation
\begin{equation}\label{eq:linNLS}
{\rm i}\, \partial_t \delta\psi =
-\tfrac12 \partial_x^2\delta\psi+\rho_0(\delta\psi+\delta\psi^*)\; ,
\end{equation}
and then $\rho(x,t)\simeq \rho_0+\sqrt{\rho_0}(\delta\psi+\delta\psi^{*})$.
In the case illustrated in Fig.~\ref{fig1}, the initial profile has,
at its maximum, a weak 15\% density increase with respect to the
background. The initial splitting of the bump if correctly described
by the linearized approach, but after the wave breaking time the
linearized evolution goes on predicting a roughly global displacement
of the two humps at constant velocity (with additional small
dispersive corrections) and clearly fails to reproduce both the
formation of DSWs and the stretching of the dispersionless part of
the profile (which reaches a quasi-triangular shape).

\section{The dispersionless stage of
  evolution}\label{DSE}

In view of the shortcomings of the linearized approximation illustrated
in Fig.~\ref{fig1}, we include nonlinear effects at all stages of the
dynamical study of the model.  By means of the Madelung substitution
\begin{equation}\label{3-2}
    \psi(x,t)=\sqrt{\rho(x,t)}\exp\left({\rm i}
\int^x u(x',t)\, dx'\right)
\end{equation}
the NLS equation \eqref{eq:nls} can be cast into a hydrodynamic-like
form for the density $\rho(x,t)$ and the flow velocity $u(x,t)$:
\begin{equation}\label{3-3}
    \begin{split}
    \rho_t+(\rho u)_{x}=0\; ,\\
    u_t+u u_{x}+\rho_{x}+\left(\frac{\rho_{x}^2}{8\rho^2}
    -\frac{\rho_{xx}}{4\rho}\right)_{x}=0 \; .
    \end{split}
\end{equation}
These equations are to be solved with the initial conditions
\eqref{rho_init} and $u(x,0)=0$.

The last term of the left hand-side of the second of Eqs.~(\ref{3-3})
accounts for the dispersive character of the fluid of light. In the
first stage of spreading of the bump, if the density gradients of
  the initial density are weak (i.e., if
  $x_0\gg \mathrm{max}\{\rho_0^{-1/2},\rho_1^{-1/2}\}$), the effects
of dispersion can be neglected, and the system (\ref{3-3}) then
simplifies to
\begin{equation}\label{3-3a}
\rho_t+(\rho u)_x=0 \; , \quad u_t +u u_x + \rho_x=0 \; .
\end{equation}
These equations can be written in a more symmetric form by introducing the
Riemann invariants
\begin{equation}\label{3-3b}
\lambda^{\pm}(x,t)=\frac{u(x,t)}{2}\pm\sqrt{\rho(x,t)} \; ,
\end{equation}
which evolve according to the system [equi\-va\-lent to
(\ref{3-3a})]:
\begin{equation}\label{3-3c1}
\partial_t\lambda^\pm +v_\pm(\lambda^-,\lambda^+)\,
\partial_x\lambda^\pm=0 \; ,
\end{equation}
with
\begin{equation}\label{3-3c2}
  v_\pm(\lambda^-,\lambda^+)
  =\tfrac{1}{2}(3\lambda^\pm+\lambda^\mp)=u\pm\sqrt{\rho} \; .
\end{equation}
The Riemann velocities \eqref{3-3c2} have a simple physical
interpretation for a smooth velocity and density distribution: $v_+$
($v_-$) corresponds to a signal which propagates downstream (upstream)
at the local velocity of sound $c=\sqrt{\rho}$ and which is dragged by
the background flow $u$.

The system \eqref{3-3c1} can be linearized by means of the hodograph
transform (see, e.g., Ref.~\cite{kamch-2000}) which consists in
considering $x$ and $t$ as functions of $\lambda^+$ and
$\lambda^-$. One readily obtains
\begin{equation}\label{eq:hodo1}
\partial_\pm x -v_\mp \partial_\pm
t=0,
\end{equation}
where $\partial_{\pm} \equiv \partial/ \partial \lambda^{\pm}$.
One introduces two auxiliary (yet unknown) functions
$W_{\pm}(\lambda^+,\lambda^-)$ such that
\begin{equation}
x - v_\pm(\lambda^-,\lambda^+)\,t = W_\pm(\lambda^-,\lambda^+).
\label{hodograph}
\end{equation}
Inserting the above expressions in \eqref{eq:hodo1} shows that the
$W^{\pm}$'s are solution of
Tsarev equations \cite{Tsa91}
\begin{equation}
\begin{split}
& \frac{\partial_-  W_+ }{ W_+ -  W_-} = \frac{\partial_- v_+}{v_+-v_-}, \\
& \frac{\partial_+  W_- }{ W_+ -  W_-} = \frac{\partial_+ v_-}{v_+-v_-}.
\end{split}
\label{Tsarev}
\end{equation}
From Eqs.~\eqref{3-3c2} and (\ref{Tsarev}) one can verify that $\partial_-
W_+ = \partial_+ W_-$, which shows that $W_+$ and $W_-$ can be sought
in the form
\begin{equation}\label{eq:pot}
  W_\pm = \partial_\pm \chi,
\end{equation}
where $\chi(\lambda^-,\lambda^+)$ plays the role of a
potential. Substituting expressions \eqref{eq:pot} in one of the Tsarev
equations shows that $\chi$ is a solution of the Euler-Poisson equation
\begin{equation}
\frac{\partial^2 \chi}{\partial \lambda^+ \partial\lambda^-} -
\frac{1}{2\,(\lambda^+ - \lambda^-)}
\left( \frac{\partial \chi}{\partial \lambda^+} -
\frac{\partial \chi}{\partial \lambda^-} \right) = 0,
\end{equation}
which can be written under the standard form
\begin{equation}\label{eq:EP}
\frac{\partial^2 \chi}{\partial \lambda^+\partial \lambda^-}
+ a(\lambda^-,\lambda^+) \frac{\partial \chi}{\partial \lambda^+} +
b(\lambda^-,\lambda^+) \frac{\partial \chi}{\partial \lambda^-} = 0,
\end{equation}
with
\begin{equation}
a(\lambda^-,\lambda^+) = - b(\lambda^-,\lambda^+) =
- \frac{1}{2\,(\lambda^+ - \lambda^-)}.
\end{equation}

\subsection{Solution of the Euler-Poisson equation}\label{sec.EP}

One can use Riemann's method to solve Eq.~\eqref{eq:EP} in the
($\lambda^+$, $\lambda^-$)--plane which we denote below as the
``characteristic plane''. We follow here the
procedure exposed in Refs.~\cite{Lud52,For2009} which applies to
non-monotonous initial distributions, such as the one corresponding to
Eq.~\eqref{rho_init}.
\begin{figure}[h!]
\centering
\includegraphics[width=\linewidth]{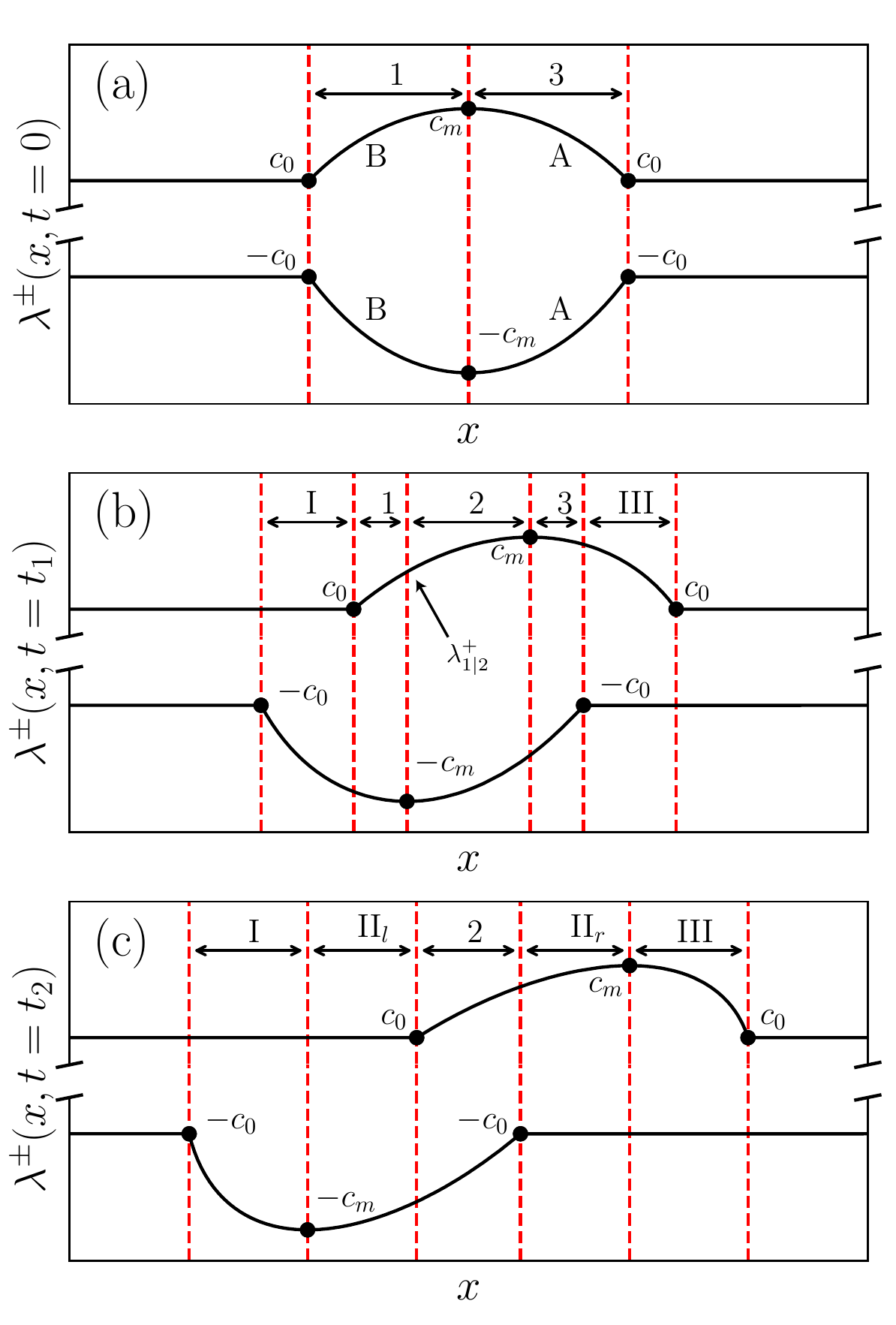}
\caption{Sketch of the distributions $\lambda^\pm(x,t)$ at several
  times. In each panel the upper solid curve represent $\lambda^+$
  (always larger than $c_0$), and the lower one $\lambda^-$ (always
  lower than $-c_0$), both plotted as functions of $x$.  Panel (a)
  corresponds to the initial distribution, in which part B corresponds
  to region 1 and part A to region 3 (see the text).  Two subsequent
  relevant stages of evolution are represented in panels (b) and (c). 
  They correspond to times $t_1<t_{ \rm \sss SW}(c_m)<t_2$, where $t_{ \rm \sss SW}(c_m)$ is 
defined in Sec. \ref{sec.match} (see also Fig. \ref{fig.char-traj}).
  For $t>0$, $\lambda^+$ ($\lambda^-$) moves to the right (to the
  left) and part B of $\lambda^+$ starts to overlap with part A of
  $\lambda^-$. This behavior initially leads to the configuration
  represented in panel (b) where a new region (labeled region 2) has
  appeared. For later convenience, we spot in this panel the value
  $\lambda^+_{1|2}(t_1)$ of the Riemann invariant $\lambda^+$ at the
  boundary between regions 1 and 2 (see the discussion in
  Sec. \ref{sec.num}). For longer time (panel (c)), region 2 remains
  while regions 1 and 3 vanish and new simple wave regions II$_l$ and
  II$_r$ appear. At even larger times (not represented), region 2 also
  vanishes and only simple-wave regions remain: the initial pulse has
  split in two simple-wave pulses propagating in opposite
  directions.}
\label{fig2}
\end{figure}

We first schematically depict in Fig.~\ref{fig2} the initial spatial
distributions $\lambda^\pm(x,0)$ of the Riemann invariants (upper
panel), and their later typical time evolution (lower panels).  We
introduce notations for some {\bf special} initial values of the Riemann
invariants: $\lambda^{\pm}(-x_0,0) = \lambda^\pm(x_0,0)=
\pm\sqrt{\rho_0} = \pm c_0$ and $\lambda^\pm(0,0)= \pm
\sqrt{\rho_m}=\pm c_m$. We also define as part A (part B) the
branch of the distribution of the $\lambda^\pm$'s which is at the
right (at the left) of the extremum. All these notations are
summarized in Fig.~\ref{fig2}(a).

At a given time, the $x$ axis can be considered as divided in five
domains, each requiring a specific treatment. Each region is
characterized by the behavior of the Riemann invariant and is
identified in the two lower panels of Fig.~\ref{fig2}. The domains in
which both Riemann invariants depend on position are labeled by arabic
numbers, the ones in which only one Riemann invariant depends on $x$
are labeled by capital roman numbers.  For instance, in region III,
$\lambda^+$ is a decreasing function of $x$ while $\lambda^-=-c_0$ is
a constant; in region 3, $\lambda^+$ is decreasing while $\lambda^-$
is increasing; in region 2 both are increasing, etc.

\begin{figure}
\includegraphics[width=\linewidth]{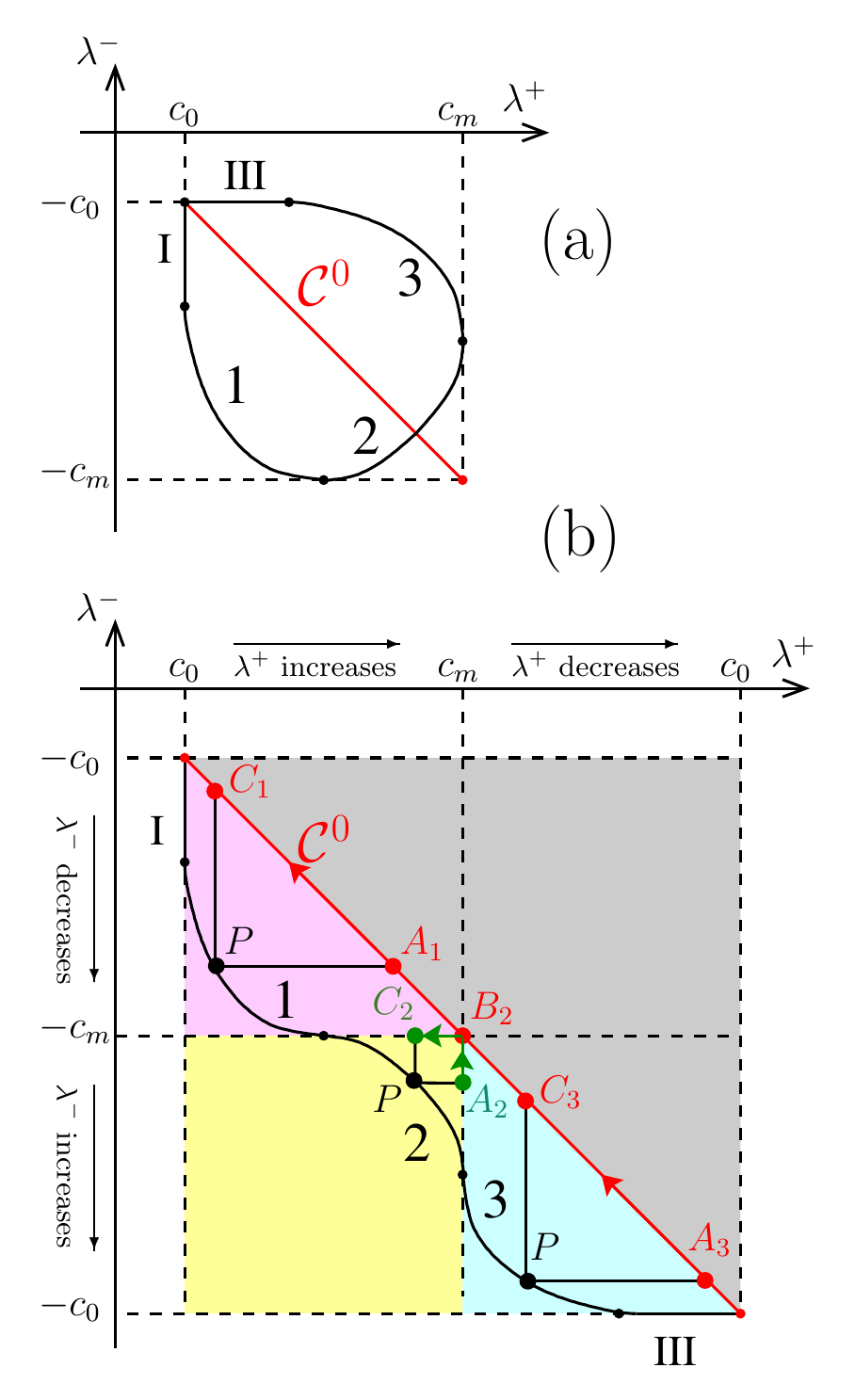}
\caption{(a) Behavior of the Riemann invariants in the characteristic
  plane at a given time $t$. (b) The same curve on the four-sheeted
  unfolded surface. The red curve $\mathcal{C}^{\, 0}$ corresponds to
  the initial condition [$\lambda^-(x,0) = - \lambda^+(x,0)$]. At
  later time, the relation between $\lambda^+(x,t)$ and
  $\lambda^-(x,t)$ is given by the black solid curve which is denoted
  as $\mathcal{C}^{\, t}$ in the main text. A generic point $P$ of
  $\mathcal{C}^{\, t}$ has coordinates ($\lambda^+$, $\lambda^-$) and
  points $C_1,A_1,B_2,C_3,A_3$ lie on the initial curve
  $\mathcal{C}^{\,0}$. Points $A_2$ and $C_2$ lie on a boundary
  between two regions. The arrows indicate the direction of
  integration in Eq. \eqref{Riemann1} and \eqref{region2}. In our
  problem, the whole gray shaded domain above ${\cal C}^0$ is
  unreachable.}\label{fig.char-plan}
\end{figure}

The values of the Riemann invariants at time corresponding to the
central panel of Fig.~\eqref{fig2} are represented in the
characteristic plane in Fig.~\ref{fig.char-plan}(a). In this plot the
straight solid lines correspond to the simple wave regions (I and III)
while the curvy lines corresponds to regions where both Riemann
invariants depend on position: the domains 1, 2 and 3. In each of
these three domains the solution $\chi$ of the Euler-Poisson equation
has a different expression. In order to describe these three branches,
following Ludford~\cite{Lud52}, we introduce several sheets in the
characteristic plane by unfolding the domain $[c_0,c_m]\times
[-c_m,-c_0]$ into a four times larger region as illustrated in
Fig.~\ref{fig.char-plan}(b). The potential $\chi(\lambda^-,\lambda^+)$
can now take a different form in each of the regions labeled 1, 2 and
3 in Fig.~\ref{fig.char-plan}(b) and still be considered as
single-valued.

We consider a flow where initially $u(x,0) = 0$, this implies that
$\lambda^+(x,0) = - \lambda^-(x,0)$.  This condition defines the curve
of initial conditions of our problem in the characteristic plane. It
is represented by a red solid line denoted as $\mathcal{C}^{\, 0}$ in
Fig.~\ref{fig.char-plan}.  We remark here that the whole region above
$\mathcal{C}^{\, 0}$ --- shaded in Fig.~\ref{fig.char-plan}(b) --- is
unreachable for the initial distribution we consider: for instance the
upper shaded triangle in region 1 would correspond to a configuration
in which $\lambda^+_{{\rm region} 1}(x,t) > |\lambda^-_{{\rm region}
  1}(x,t)|$, which does not occur in our case, see Fig.~\ref{fig2}(b).

Before establishing the expression for $\chi$ is the three relevant
regions of Fig.~\ref{fig.char-plan} it is convenient to define the
inverse functions of the initial $\lambda$ profiles in both parts A
and B of Fig.~\ref{fig2}(a). The symmetry of the initial conditions
makes it possible to use the same functions for $\lambda^+ \in
[c_0,c_m\,]$ and for $\lambda^- \in [-c_m,-c_0\,]$:
\begin{equation}\label{InvFunc}
\begin{cases}
x = w^{\sss\rm A}(\lambda^\pm)=x_0
\sqrt{1-
\frac{\displaystyle (\lambda^\pm)^2-\rho_{0}}{\displaystyle \rho_{m}-\rho_{0}}}
\quad \text{if} \;\; x>0, \\[2mm]
x= w^{\sss\rm B}(\lambda^\pm)=-x_0
\sqrt{1-
\frac{\displaystyle(\lambda^\pm)^2-\rho_{0}}{\displaystyle\rho_{m}-\rho_{0}}}
\quad \text{if} \;\; x<0.
\end{cases}
\end{equation}
For $t=0$, using Eqs.~\eqref{hodograph} and \eqref{eq:pot},
the boundary conditions read
\begin{equation}\label{eq18}
\left. \frac{\partial \chi}{\partial \lambda^{\pm}}
\right|_{\lambda^{\pm}(x,t=0)} = x =
w^{{\rm\sss A}/{\rm\sss B}}(\lambda^\pm),
\end{equation}
where the superscript B holds in region 1 (when $x<0$)
and A holds in region 3 ($x>0$).
Formula \eqref{eq18} requires some explanation: its left-hand side is a
function of two variables $\lambda^+$ and $\lambda^-$ which is
evaluated for $\lambda^-=-\lambda^+$; its right-hand side is expressed
by the same function in terms of $\lambda^+$ or $\lambda^-$ since the
function $w^{{\rm\sss A}}$ and $w^{{\rm\sss B}}$ depend only on the
square of their argument.  The boundary conditions \eqref{eq18}
corresponds to a potential $\chi$ which takes the following form along
${\cal C}^{\, 0}$:
\begin{equation}
\begin{split}
\chi^{(n)}(\lambda^-=-\lambda^+,\lambda^+)=&
\int_{c_0}^{\lambda^+}\!\!\!\! w^{{\rm\sss A}/{\rm\sss B}}(r) \, dr\\
& + \int_{-c_0}^{\lambda^-}\!\!\!\! w^{{\rm\sss A}/{\rm\sss B}}(r) \, dr,
\end{split}
\end{equation}
where $n=1$ or 3 and, in the right hand side, and the superscript A
(B) holds when $n=3$ ($n=1$).  For the specific initial condition we
consider ($u(x,0)\equiv 0$ and $\rho(x,0)$ an even function of $x$),
$w^{{\rm\sss A}}$ and $w^{{\rm\sss B}}$ are even functions and thus
our choice of integration constants yields $\chi= 0$ along
${\cal C}^{\, 0}$.

Let us now consider a point $P$, lying either in region 1 or 3 (the
case of region 2 is considered later), with coordinates
$(\lambda^+,\lambda^-)$ in the characteristic plane. We introduce
points $A_1$, $A_3$, $C_1$ and $C_3$ which are located on the curve
$\mathcal{C}^{\, 0}$, with geometrical definitions obvious from
Fig.~\ref{fig.char-plan}(b). Note the different subscripts for $C$ and
$A$: subscript 1 (3) is to be used if $P$ is in region 1 (3). One can
obtain the value of $\chi$ at the point $P$ from Riemann's method
(see, e.g. Ref.~\cite{Som64}), the general solution reads
\begin{equation}
\begin{split}
\chi^{(n)}(P) = & \frac{1}{2} \chi(C_{n}) R(C_{n}) +
\frac{1}{2} \chi(A_{n}) R(A_{n}) \\
& - \int_{A_{n}}^{C_{n}} V \, dr + U \, ds,
\end{split}
\label{Riemann1}
\end{equation}
with
\begin{equation}\label{Riemann1a}
\begin{split}
& U(s,r) = \frac{1}{2} \left(R\, \frac{\partial \chi}{\partial s}
- \chi\, \frac{\partial R}{\partial s} \right)
+ a R\, \chi,\\
& V(s,r) = \frac{1}{2} \left(\chi\, \frac{\partial R}{\partial r}
- R\, \frac{\partial \chi}{\partial r} \right) - b R\, \chi,
\end{split}
\end{equation}
where
\begin{equation}
R( s, r ) =
\frac{2}{\pi} \sqrt{\frac{r - s}{\lambda^+ - \lambda^-}} \,
\mathrm{K}[m(s,r)] ,
\end{equation}
$\mathrm{K}$ being the complete elliptic integral of the first
kind and
\begin{equation}
m(s,r ) =
\frac{(\lambda^+-r)\,(\lambda^--s)}
{(r - s) \,(\lambda^+-\lambda^-)}.
\label{Riemann2}
\end{equation}
is the associated parameter (we follow here the convention of
  Ref. \cite{AS}).  In our case, the symmetries of the initial profile lead
to many simplifications in the above formulas \eqref{Riemann1} and
\eqref{Riemann1a}. Along the curve $\mathcal{C}^{\, 0}$ one has
$\chi=0$. This implies that
$\chi^{(n)}(A_{n}) = \chi^{(n)}(C_{n}) =0$, and along the integration
path going from $A_{n}$ to $C_{n}$ one has
\begin{equation}
U = \tfrac{1}{2} \,w^{{\rm\sss A}/{\rm\sss B}}(r) \,
R(s=-r,r) = -V\; ,
\end{equation}
where the superscript A (B) holds when $P$ is in region 3 (region 1).
Explicit evaluation of expression \eqref{Riemann1} then yields
\begin{equation}
\chi^{(n)}(P) = \frac{2\,\sqrt{2}}{\pi \,\sqrt{\lambda^+-\lambda^-}}
\int_{-\lambda^-}^{\lambda^+} \!\!\!\! \sqrt{r}\,
\mathrm{K}[m(r)]\,w^{{\rm\sss A}/{\rm\sss B}}(r)\,dr,
\label{Region13}
\end{equation}
where
\begin{equation}\label{mder}
m(r)\equiv m(-r,r)=
\frac{(\lambda^+-r)\,(\lambda^-+r)}{2\,r \,(\lambda^+-\lambda^-)}\; .
\end{equation}
To calculate $\chi(P)$ in region 2 we define three points: $A_2$,
$B_2$ and $C_2$, see Fig.~\ref{fig.char-plan}(b). Point $B_2$ is on
the curve $\mathcal{C}^{\, 0}$, at the junction between regions 1, 2
and 3. Point $A_2$ lies on the characteristic curve $\lambda^+ = c_m$,
on the boundary between regions 2 and 3, whereas point $C_2$ lies on
the characteristic $\lambda^- = -c_m$, on the boundary between regions
1 and 2.  Then, from Eqs.~\eqref{Riemann1} to \eqref{Riemann2}, one
can easily find that in region 2
\begin{equation}
\begin{split}
\chi^{(2)}(P) & = \chi(B_2)\, R(B_2)
+ \int_{B_2}^{C_2} \left( \frac{\partial \chi}{\partial r}
+ b \chi\right) \, R_1(r) \, dr \\
& - \int_{A_2}^{B_2}
\left( \frac{\partial  \chi}{\partial s} + a \chi\right)
\, R_2(s) \, ds,
\end{split}
\label{region2}
\end{equation}
where
\begin{equation}
\begin{cases}
&R_1(r) \equiv  \frac{\displaystyle 2}{\displaystyle \pi}
\sqrt{\displaystyle\frac{r + c_m}{\lambda^+ - \lambda^-}} \,
\mathrm{K}[m_1(r)],   \\[2mm]
& m_1(r) =
\frac{\displaystyle (r - \lambda^+)\,( c_m +\lambda^-)}
{\displaystyle(r- \lambda^-) \,(\lambda^++c_m)},
\end{cases}
\end{equation}
and
\begin{equation}
\begin{cases}
& R_2(s) =
\frac{\displaystyle 2}{\displaystyle \pi}
\sqrt{\displaystyle\frac{c_m - s}{\lambda^+ - \lambda^-}}
\mathrm{K}[m_2(s)], \\[2mm]
& m_2(s) =
\frac{\displaystyle(c_m - \lambda^+)\,( \lambda^- - s)}
{\displaystyle (c_m - \lambda^-) \,(\lambda^+- s)}.
\end{cases}
\end{equation}
Note that in formula \eqref{region2} one has $\chi(B_2)=0$ and
the value of $\chi$ along the
integration lines $B_2C_2$ and $A_2B_2$ is known from the previous
result \eqref{Region13}. After some computation we eventually get the
following expression for $\chi(P)$ in region 2:
\begin{equation}\label{eq:chi}
\begin{split}
\chi^{(2)}(P) & = \frac{2\,\sqrt{2}}{\pi \,\sqrt{\lambda^+-\lambda^-}}
\left[ \int_{c_m}^{\lambda^+} \!\!\!\!\!\sqrt{r}\,
\mathrm{K}[m_0(r\,;\lambda^+)]\,w^{\rm\sss B}(r)\,dr \right. \\
& \left. +  \int_{-\lambda^-}^{c_m}
\sqrt{r}\,\mathrm{K}[m_0(r\,;-\lambda^-)]\,w^{\rm\sss A}(r)\,dr \right] \\
&+  \frac{4\,\sqrt{2}}{\pi^2 \,\sqrt{\lambda^+-\lambda^-}}
\left[ \int_{c_m}^{\lambda^+} \sqrt{r}
\,w^{\rm\sss B}(r)\, f_1(r) \, dr \right. \\
&\left. + \int_{-\lambda^-}^{c_m}
\sqrt{r} \,w^{\rm\sss A}(r)\, f_2(r ) \, dr \right],
\end{split}
\end{equation}
where we have introduced the notations
\begin{equation}
\begin{split}
f_{1}(r) & = \int_{\lambda^+}^{r} \mathrm{K}[m_0(r\, ; u )] \,
\frac{\partial \mathrm{K}[m_{1}(u)]}{\partial u} \,  du \; ,\\
f_{2}(r) & = \int_{-\lambda^-}^{r} \mathrm{K}[m_0(r\, ; u )] \,
\frac{\partial \mathrm{K}[m_{2}(- u)]}{\partial u} \,  du \; ,\end{split}
\end{equation}
with
\begin{equation}
m_0(r;u)=\frac{(r-u)(c_m-r)}{2 r (u+c_m)}\; .
\end{equation}
In many instances one can actually simplify the above expressions
\eqref{Region13} and \eqref{eq:chi}: for
reasonable values of $c_m$ (chosen to be of same order as $c_0$ in our
simulations) the elliptic integral $\mathrm{K}(m)$ turns out to be
approximately equal to $\pi/2$ for all points $P$ in the three
regions. In this case, the exact expressions \eqref{Region13} and
\eqref{eq:chi} can be
replaced by a simple approximation $\chi(P)\simeq {\chi}_{\rm app}(P)$
which reads, when $P$ is in region $n=1$ or $3$:
\begin{equation}
\chi^{(n)}_{\rm app}(\la_-,\la_+) =
\frac{\sqrt{2}}{\sqrt{\lambda^+-\lambda^-}}
\int_{-\lambda^-}^{\lambda^+} \!\!\!\! \sqrt{r}\,
\,w^{{\rm\sss A}/{\rm\sss B}}(r)\,dr,
\label{Region13app}
\end{equation}
where the superscript A (B) holds when $n=3$ ($n=1$). When $P$ is in region
$2$ one gets:
\begin{equation}\label{chi2}
\begin{split}
{\chi}^{(2)}_{\rm app}(\la_-,\la_+) & = \frac{\sqrt{2}}{\sqrt{\lambda^+-\lambda^-}}
\int_{c_m}^{\lambda^+} \sqrt{r}\,w^{\rm\sss B}(r)\,dr
\\
& + \frac{\sqrt{2}}{\sqrt{\lambda^+-\lambda^-}} \int_{-\lambda^-}^{c_m}
\sqrt{r}\,w^{\rm\sss A}(r)\,dr.
\end{split}
\end{equation}
This approximation greatly simplifies the numerical determination of
the integrals involved in the solution of the problem. We have checked
that it is very accurate in all the configurations we study in the
present work. The reason for its validity is easy to understand in
regions 1 and 3: the argument of the elliptic integral K in
Eq. \eqref{Region13} is zero at the two boundaries of the integration
domain ($r=-\lambda^-$ and $r=\lambda^+$) and reaches a maximum when
$r=\sqrt{-\lambda^-\lambda^+}$, taking the value
\begin{equation}\label{mmax}
0\le m_{\rm max}=\frac{1}{2}\left(
1 -\frac{2\, \sqrt{-\lambda^-\lambda^+}}{\lambda^+-\lambda^-}\right)\le
\frac{1}{2}\; .
\end{equation}
As time varies, the largest value of of $m_{\rm max}$ is reached at
the point where region 3 disappears, when $\lambda^+=c_m$ and
$\lambda^-=-c_0$.  For $c_m/c_0\sim 1$ this value is typically much
lower than the upper bound $1/2$ of Eq. \eqref{mmax}. For instance, in
the numerical simulations below, we take $\rho_0=0.5$ and $\rho_m=2$
one gets accordingly $c_0=\sqrt{0.5}$ and $c_m=\sqrt{2}$ and the
corresponding largest value of $m_{\rm max}$ is about
$\simeq 2.9\times 10^{-2}$.

\subsection{Simple wave regions}\label{sec.match}

\begin{figure}
\includegraphics[width=\linewidth]{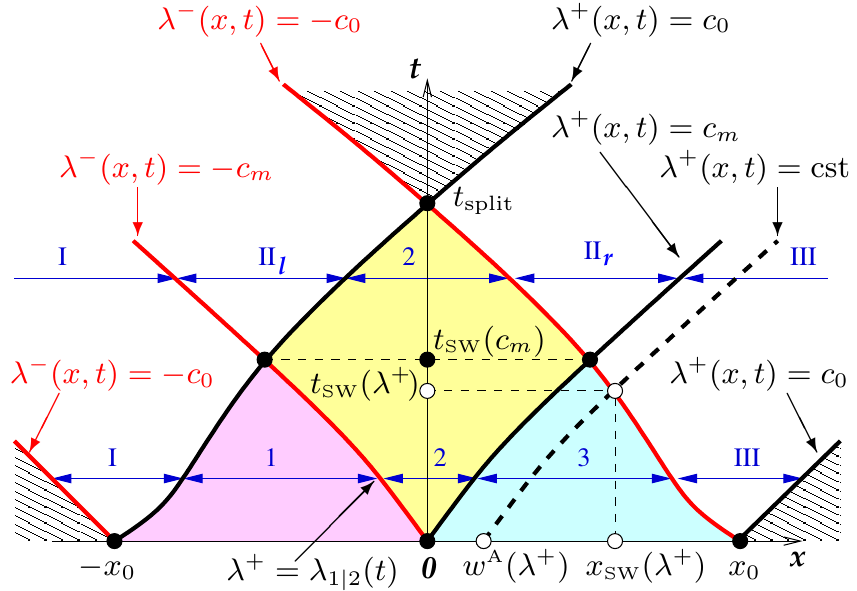}
\caption{Sketch of the characteristics in the $(x,t)$ plane. The black
  (red) solid lines are specific characteristics for $\lambda^+$
  ($\lambda^-$) stemming from the edges of the hump and from its
  maximum.  The thick dashed line is a generic characteristics for
  $\lambda^+$. In the hatched regions both Riemann invariants are
  constant ($\lambda^{\pm}(x,t)=\pm c_0$) and the profile is flat. In
  the colored regions both Riemann invariants depend on position (the
  color code is the same as in Fig.~\ref{fig.char-plan}: region 1
    is pink, region 2 is yellow, and region 3 is cyan). In the white
  regions only one Riemann invariant depends on position: one has a
  simple wave. The different notations are explained in the main
  text.}
\label{fig.char-traj}
\end{figure}

Once $\chi$ has been computed in the domains 1, 2 and 3 where two
Riemann invariants depend on position, it remains to determine the
value of $\lambda^+$ and $\lambda^-$ in the simple wave regions. Let
us for instance focus on region III, in which $\lambda^-=-c_0$ and
$\lambda^+$ depends on $x$ and $t$. The behavior of the
characteristics in the $(x,t)$ plane is sketched in
Fig.~\ref{fig.char-traj}.
One sees in this figure that the characteristic of a given value of
$\lambda^+$ enters the simple wave region III at a given time which we
denote as $t_{\rm\sss SW}(\lambda^+)$ and a given position $x_{\rm\sss
  SW}(\lambda^+)$. Beyond this point the
characteristic becomes a straight line and the general solution of
Eq. \eqref{3-3c1} for $\lambda^+$ is known to be of the form
\begin{equation}\label{IIIa}
x-v_+(-c_0,\lambda^+) t=h(\lambda^+)\; ,
\end{equation}
where the unknown function $h$ is determined by boundary conditions.
From Eq.~\eqref{hodograph} one sees that just at the
boundary between regions 3 and III one has
\begin{equation}\label{IIIb}
x_{\rm\sss SW}(\lambda^+)
-v_+(-c_0,\lambda^+)t_{\rm\sss SW}(\lambda^+)=W_+^{(3)}(-c_0,\lambda^+)\; ,
\end{equation}
where $W_+^{(3)}=\partial_+\chi^{(3)}$. This shows that in
Eq. \eqref{IIIa} the unknown function $h(\lambda^+)$ is equal to
$W_+^{(3)}(-c_0,\lambda^+)$. The equation of the characteristic in
region III thus reads
\begin{equation}\label{IIIc}
x-v_+(-c_0,\lambda^+)t=W_+^{(3)}(-c_0,\lambda^+)\; .
\end{equation}
A similar reasoning shows that in region I one has
\begin{equation}\label{IIId}
x-v_-(\lambda^-,c_0)t=W_{-}^{(1)}(\lambda^-,c_0)\; .
\end{equation}
For time larger than $t_{\rm\sss SW}(c_m)$, the regions 1 and 3
disappear and two new simple wave regions appear which we denote as
II$_l$ and II$_r$, see Fig.~\ref{fig.char-traj} and also the lower
panel of Fig.~\ref{fig2}. The same reasoning as above shows that in
these regions the characteristics are determined by
\begin{equation}\label{IIIe}
x-v_+(-c_0,\lambda^+)t= W_{+}^{(2)}(-c_0,\lambda^+)\; , \quad\mbox{in II$_r$,}
\end{equation}
and
\begin{equation}\label{IIIf}
x-v_-(\lambda^-,c_0)t= W_{-}^{(2)}(\lambda^-,c_0)\; , \quad\mbox{in II$_l$.}
\end{equation}

\subsection{Solution of the dispersionless problem and comparison with
  numerical simulations}\label{sec.num}

The problem is now solved: having determined $\chi$ in regions 1, 2
and 3 (see Sec.~\ref{sec.EP}), we obtain $W_{\pm}$ in these regions
from Eqs.~\eqref{eq:pot}.

$\bullet$ It is then particularly easy to find the values of
$\lambda^+$ and $\lambda^-$ in the simple wave regions. For instance,
in region III, one has $\lambda_-=-c_0$, and for given $x$ and $t$,
$\lambda^+$ is obtained from Eq. \eqref{IIIc}. The same procedure is
to be employed in the simple wave regions I, II$_r$ and II$_l$ where
the relevant equations are then Eqs. \eqref{IIId}, \eqref{IIIe},
\eqref{IIIf} respectively.

$\bullet$ To determine the values of $\lambda^+$ and $\lambda^-$ as
functions of $x$ and $t$ in regions 1, 2 and 3 one follows a different
procedure which is detailed below, but which essentially consists in
the following: for a given time $t$ and a given region $n$ ($n=1$, 2
or 3) one picks one of the possible values of $\lambda^+$. From Eqs.
\eqref{hodograph} $\lambda^-$ is then solution of
\begin{equation}
\frac{W_+^{(n)}(\lambda^-,\lambda^+)-W_-^{(n)}(\lambda^-,\lambda^+)}
{v_+(\lambda^-,\lambda^+) - v_-(\lambda^-,\lambda^+)} + t = 0\; ,
\label{eq_dispn}
\end{equation}
and $x$ is determined by either one of Eqs.~\eqref{hodograph}. So, for
given $t$ and $\lambda^+$ in region $n$, one has determined the values
of $\lambda^-$ and $x$. In practice, this makes it possible to
associate a couple $(\lambda^-,\lambda^+)$ in region $n$ to each
$(x,t)$.

The procedure for determining the profile in regions 1, 2, and 3 which
has just been explained has to be implemented with care, because the
relevant regions to be considered and their boundaries change with
time; for instance regions 1 and 3 disappear when
$t>t_{\sss\rm SW}(c_m)$. It would be tedious to list here all the
possible cases and we rather explain the specifics of the procedure by
means of an example: the determination of $\lambda^+$ and $\lambda^-$
in region 1 when $t<t_{\sss\rm SW}(c_m)$.

One starts by determining the value of $\lambda^+$ along the
characteristic $\lambda^-=-c_m$ at time $t$ (see
Fig.~\ref{fig.char-traj}). This value of $\lambda^+$
defines the boundary between
regions 1 and 2 and we accordingly denote it as $\lambda^+_{1|2}(t)$,
it is represented in Fig.~\ref{fig2}(b). From Eqs. \eqref{hodograph}
it is a solution of
\begin{equation}
\frac{W_+^{(1)}(-c_m,\lambda^+_{1|2})-W_-^{(1)}(-c_m,\lambda^+_{1|2})}
{v_+(-c_m,\lambda^+_{1|2}) - v_-(-c_m,\lambda^+_{1|2})} + t = 0\; .
\end{equation}
We then know that, in region 1, at time $t$, $\lambda^+$ takes all
possible values between $c_0$ and $\lambda^+_{1|2}(t)$. Having
determined the precise range of variation of $\lambda^+$ we can now,
for each possible $\lambda^+$, determine $\lambda^-$ from
Eq.~\eqref{eq_dispn} (with $n=1$) and follow the above explained
procedure.

\begin{figure}
\centering
\includegraphics[scale=0.6]{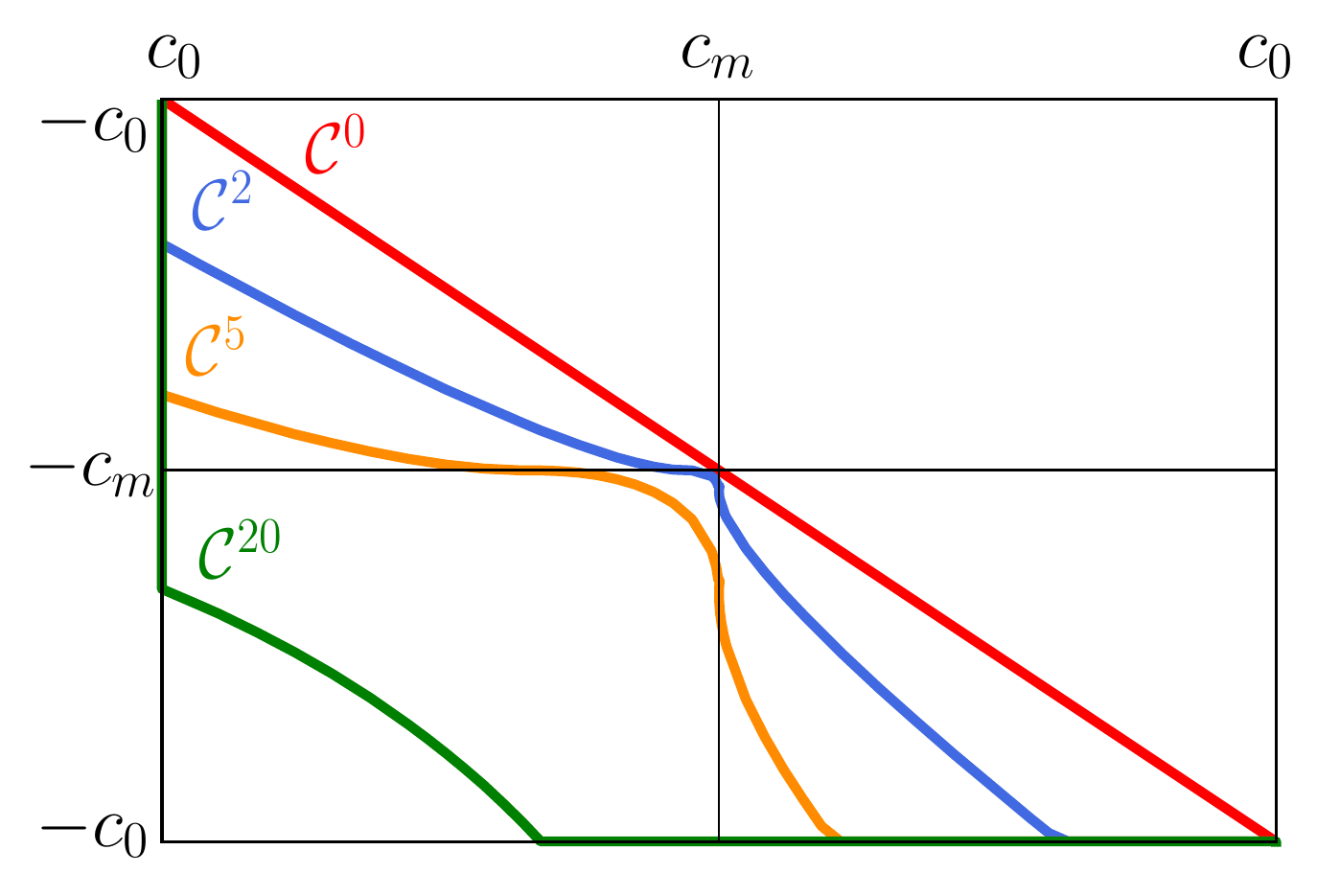}
\caption{Theoretical curves $\mathcal{C}^{\, t}$ representing
  $\lambda^-$ as a function of $\lambda^+$ at a given time in the
  characteristic plane. The curves are plotted for $t=0$
  ($\mathcal{C}^{\, 0}$, in red), $t=2$ (blue), $t=5$ (orange) and
  $t=20>t_{\rm\sss SW}(c_m)$ (green). The corresponding initial distribution
  $\la^\pm(x,0)$ is schematically represented in the upper part of
  Fig.~\ref{fig2}. We take here $c_0 = 1/\sqrt2$ and $c_m=\sqrt{2}$.}
\label{fig5}
\end{figure}

$\bullet$ The approach described in the present section makes it
possible to determine the curve ${\cal C}^{\, t}$ representing, at time
$t$, the profile in the unfolded characteristic plane. A sketch of
${\cal C}^{\, t}$ was given in Fig.~\ref{fig.char-plan}(b), it is now
precisely represented in Fig.~\eqref{fig5} for several values of $t$,
with also the initial curve ${\cal C}^0$.

\begin{figure}
\centering
\includegraphics[width=\linewidth]{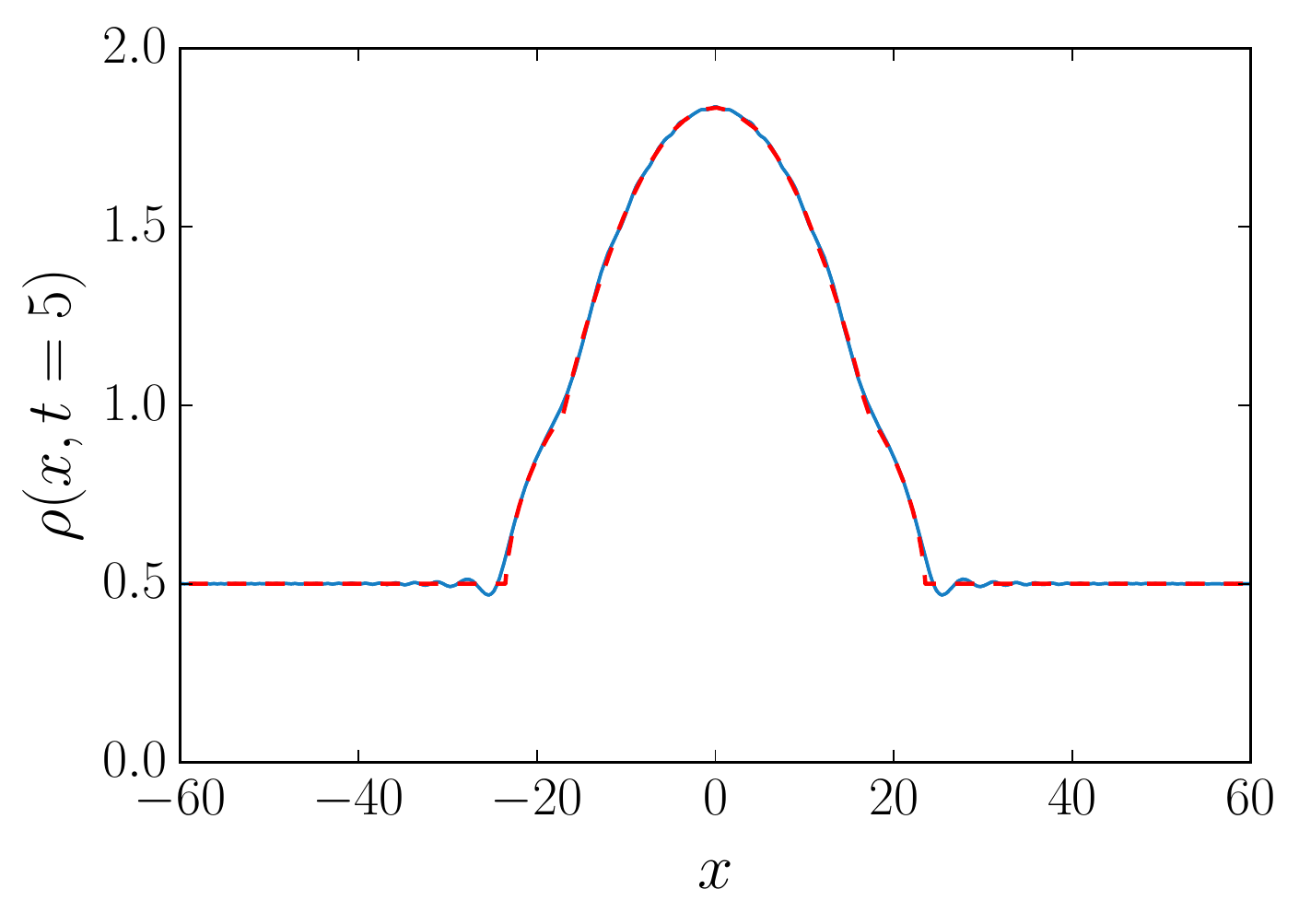}
\caption{Comparison between theory and simulations for $t=5$. The red
  dashed line is extracted from the exact solution of the
  dispersionless system \eqref{3-3c1} (see the text), while the blue
  curve displays the numerical solution of Eq.~\eqref{eq:nls} with the
  initial conditions $u(x,0)=0$ and $\rho(x,0)$ given by
  Eq.~\eqref{rho_init} taking $\rho_0=0.5$, $\rho_1=1.5$ (i.e.,
  $\rho_m=2$) and $x_0=20$. The corresponding initial distributions
  $\la^\pm(x,0)$ are drawn schematically in Fig.~\ref{fig2}(a) with here
  $c_0= \sqrt{\rho_0}=\sqrt{0.5}$ and $c_m= \sqrt{\rho_m}=\sqrt{2}$.}
\label{fig7a}
\end{figure}

\begin{figure}
\centering
\includegraphics[width=\linewidth]{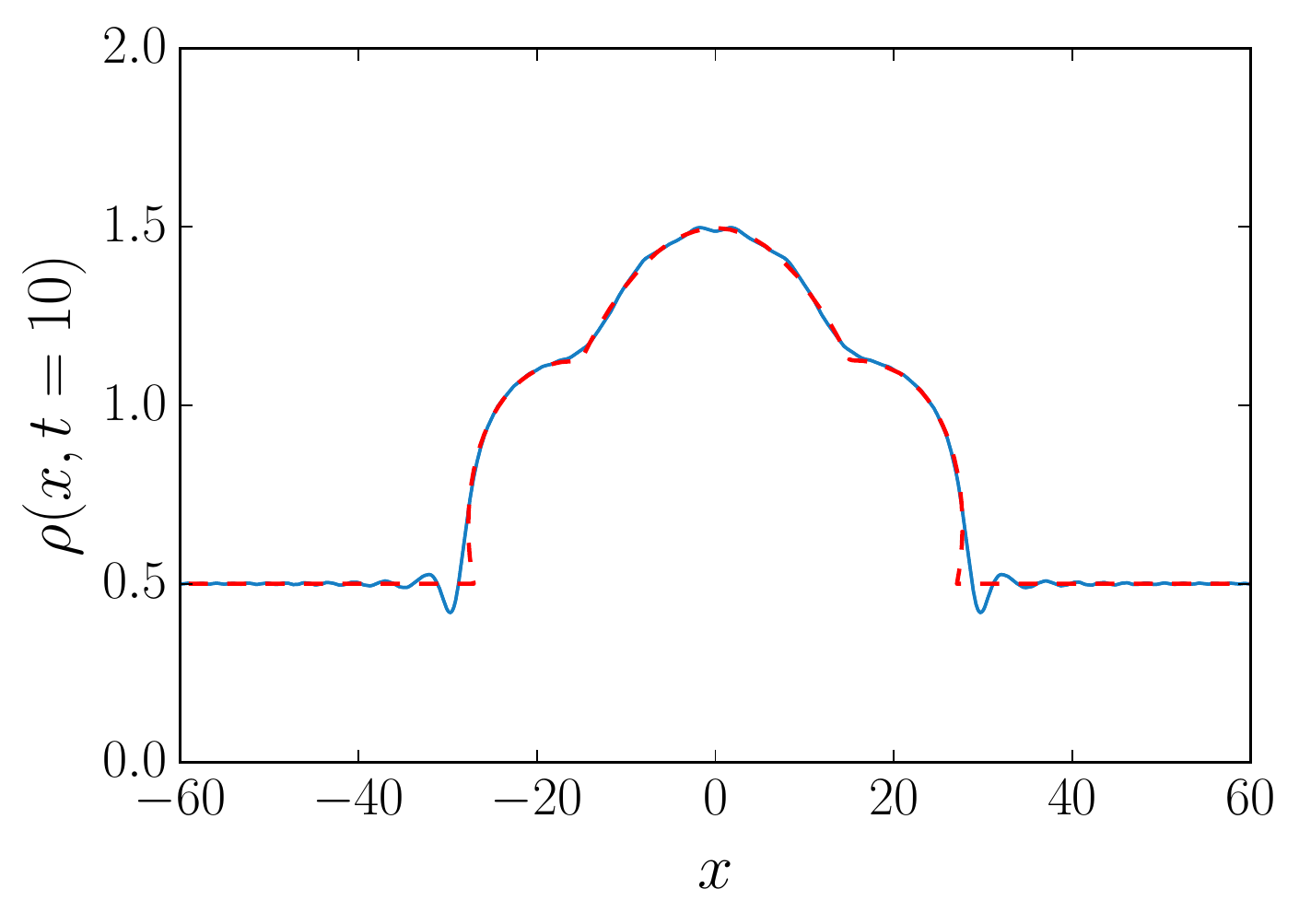}
\caption{Same as Fig.~\ref{fig7a} with here $t=10$. Notice that the
  dispersionless treatment leads to small regions of multivalued
  profile at both edges of the pulse.}
\label{fig7b}
\end{figure}

\begin{figure}
\centering
\includegraphics[width=\linewidth]{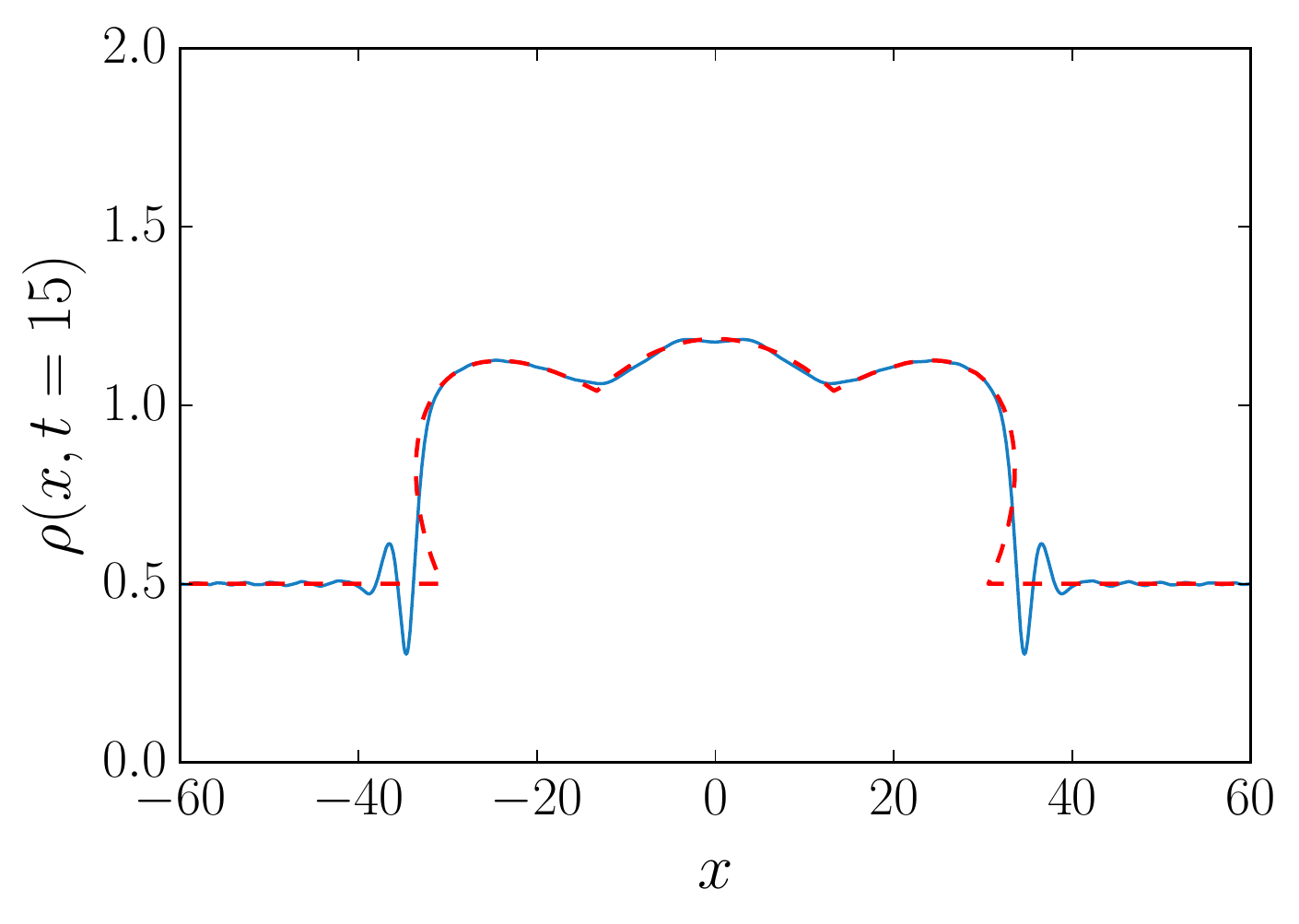}
\caption{Same as Figs.~\ref{fig7a} and \ref{fig7b} with here $t =15$.
  The multivaluedness of the theoretical profile is here obvious. It is
  associated to the formation of dispersive shocks at both edges of the
  pulse.}
\label{fig7c}
\end{figure}

Once $\lambda^+$ and $\lambda^-$ have been determined as functions of
$x$ and $t$, the density and velocity profiles are obtained through
Eqs.~\eqref{3-3b}. One obtains an excellent description of the initial
dispersionless stage of evolution of the pulse, as demonstrated by the
very good agreement between theory and numerical simulations
illustrated in Figs.~\ref{fig7a} and \ref{fig7b}. These figures,
together with Fig.~\ref{fig7c}, compare at different times the
theoretical density profile $\rho(x,t)$ with the one obtained by
numerical integration of Eq.~\eqref{eq:nls}, taking the initial
condition $u(x,0)=0$ and $\rho(x,0)$ given by \eqref{rho_init} with
$\rho_0=0.5$, $\rho_m=2$ and $x_0=20$.
A similar agreement is obtained for the velocity profile
$u(x,t)$. Note that for time $t=5$, some small diffractive
contributions at the left and right boundary of the pulse are not
accounted for by our dispersionless treatment (see
Fig.~\ref{fig7a}). At larger time, the density profile at both ends of
the pulse steepens and the amplitude of these oscillations accordingly
increases. There exists a certain time, the wave breaking time
$t_{\sss\rm WB}$, at which nonlinear spreading leads to a gradient
catastrophe; the dispersionless approximation subsequently predicts
a nonphysical multivalued profile, as can be already seen in
Fig.~\ref{fig7b} and more clearly in Fig.~\ref{fig7c}. The time
$t_{\sss\rm WB}$ can be easily computed if the wave breaking occurs at
the simple-wave edges of the pulse (see, e.g., \cite{LL-6}) as it
happens in our case, when the simple waves I and III break. These
edges propagate with the ``sound'' velocity $c_0$ over a flat
background and, at the wave breaking moment, the profile of
$\lambda^+$ in region III (or $\lambda^-$ in region I) has a vertical
tangent line in the limit $\lambda^+\to c_0$ ($\lambda^-\to-c_0$),
that is $\partial x/\partial\lambda^{\pm}\to0$ as
$\lambda^{\pm}\to\pm c_0$.  Then differentiation of the simple-wave
solutions (\ref{IIIc}) or (\ref{IIId}) gives at once
\begin{equation}\label{tWB-est0}
t_{\sss\rm WB}=\frac23\left|\frac{dW^{(3)}_+(-c_0,\lambda^+)}{d\lambda^+}
\right|_{\lambda^+=c_0}
\end{equation}
(for definiteness we consider the simple wave in the region
III). Substitution of the expression for $W^{(3)}_+(-c_0,\lambda^+)$
in the above relation yields after simple calculations \cite{foot1}
\begin{equation}\label{tWB-est}
  t_{\sss\rm WB}=\frac23\left|\frac{dw^{{\rm\sss A}}}{d\lambda^+}
  \right|_{\lambda^+=c_0}.
\end{equation}
The numerical value of $t_{\sss\rm WB}$ is equal to $\simeq 6.3$ for
our choice of initial condition, in excellent agreement with the onset
of double valuedness of the solution of the Euler-Poisson. In
dispersive nonlinear systems the wave breaking is regularized by
formation of regions with large oscillations of density and flow
velocity, whose extend increases with time. This situation is typical
for the formation of dispersive shock waves, and requires a nonlinear
treatment able to account for dispersive effects. Such an approach is
introduced in the next section, but before turning to this aspect, we
now compute an important characteristics time: the time
$t_{\rm split}$ at which the initial bump has exactly split in two
separated parts. For $t>t_{\rm split}$ a plateau of constant density
$\rho_0$ develops between the two separated humps, as illustrated for
instance in Fig. \ref{fig1}. One can see from Fig. \ref{fig.char-traj}
that $t_{\rm split}=t_{\rm\sss SW}(c_0)$ and can thus be computed from
Eqs. \eqref{hodograph} as
\begin{equation}\label{ts1}
t_{\rm split}=
\frac{W_-^{(2)}(-c_0,c_0)-W_+^{(2)}(-c_0,c_0)}{v_+(-c_0,c_0)-v_-(-c_0,c_0)}\; .
\end{equation}
In the right hand side of the above equation one has
$W_\pm^{(2)}=\partial_\pm\chi^{(2)}$ where it is legitimate to use the
expression \eqref{chi2} since one is in the limiting
case where $\lambda^+=-\lambda^-$. This yields at once
\begin{equation}\label{ts2}
t_{\rm split}=\frac{x_0}{c_0}+\frac{1}{4\, c_0^{5/2}}
\int_{c_0}^{c_m}\!\!\!\sqrt{r}\left(w^{\rm\sss A}(r)-w^{\rm\sss B}(r)\right)dr \; .
\end{equation}
In the limit of a very small initial bump, $c_m$ is very close to
$c_0$, and the second term of the right hand side of Eq.~\eqref{ts2}
is negligible. In this case a linear approach is valid: the two
sub-parts of the bump move, one to the right, the other to the left,
at velocities $\pm c_0$ and a time $t_{\rm split}\simeq x_0/c_0$ is
needed for their complete separation. The second term of the right
hand side of Eq.~\eqref{ts2} describes the nonlinear correction to this
result. For the initial profile \eqref{rho_init} the expressions of
$w^{\sss\rm A}$ and $w^{\sss\rm B}$ are given in Eq.~\eqref{InvFunc} and
one directly obtains from Eq.~\eqref{ts2}:
\begin{equation}\label{ts3}
t_{\rm split}=\frac{x_0}{c_0}\Big(1+G(\rho_1/\rho_0) \Big) \; ,
\end{equation}
where
\begin{equation}\label{ts4}
G(X)=\frac{X}{4} \int_0^{1} \frac{\sqrt{1-u}}{(1+X u)^{1/4}} du \; .
\end{equation}
In the simulations, we took $x_0=20$, $c_0=\sqrt{0.5}$,
$\rho_1/\rho_0=3$ and formula \eqref{ts3} then yields
$t_{\rm split}\simeq 40.1$. Note that in this case the simple linear
estimate would be $x_0/c_0\simeq 28.3$.  The accuracy of the result
\eqref{ts3} can be checked against numerical simulations, by plotting
the numerically determined central density of the hump $\rho(x=0,t)$
as a function of time and checking that it just reaches the background
value at $t=t_{\rm split}$. This is indeed the case: for the case we
consider here $\rho(x=0,t=40.1)$ departs from $\rho_0$ by only 3 \%.

For a small bump with $\rho_1\ll \rho_0$, the weak nonlinear correction to
the linear result is obtained by evaluating the small $X$ behavior of
the function $G$ in \eqref{ts4}. This yields
\begin{equation}\label{ts5}
t_{\rm split}\simeq \frac{x_0}{c_0}
\left(1+\frac{1}{6} \,
\frac{\rho_1}{\rho_0} - \frac{1}{60} \,
\left(\frac{\rho_1}{\rho_0}\right)^2 + \cdots
 \right) \; .
\end{equation}
For the numerical values for which we performed the simulations,
stopping expansion \eqref{ts5} at first order in $\rho_1/\rho_0$
yields $t_{\rm split}\simeq 42.4$. At next order one gets
$t_{\rm split}\simeq 38.2$.  These values are reasonable upper and
lower bounds for the exact result. Of course, the expansion is more
efficient for lower values of $\rho_1/\rho_0$: even for the relatively
large value $\rho_1/\rho_0=1$, expansion \eqref{ts5} gives an estimate
which is off the exact result \eqref{ts3} by only 0.3 \%.

\section{Whitham theory and the generalized hodograph
  method}\label{WGH}

In this section we first give a general presentation of
Whitham modulational theory (Sec.~\ref{sec:DSW}) and then discuss
specific features of its implementation for the case in
which we are interested (Sec.~\ref{sec:GHM}).

\subsection{Periodic solutions and their modulations}\label{sec:DSW}

The NLS equation \eqref{eq:nls} is equivalent to the system
\eqref{3-3} which admits nonlinear periodic solutions that can be
written in terms of four parameters
$\lambda_1\le \lambda_2\le \lambda_3\le \lambda_4$ in the form
(see, e.g., Ref. \cite{kamch-2000})
\begin{equation}\label{4-2}
\begin{split}
    \rho(x,t)= &\tfrac14(\lambda_4-\lambda_3-\lambda_2+\lambda_1)^2+\\
 &(\lambda_4-\lambda_3)(\lambda_2-\lambda_1)\times \\
     &\mathrm{sn}^2
\left(\sqrt{(\lambda_4-\lambda_2)(\lambda_3-\lambda_1)}\,(x-Vt),m\right) \; ,\\
  u(x,t)=& V - \frac{C}{\rho(x,t)}\; ,\end{split}
\end{equation}
where $\mathrm{sn}$ is the Jacobi elliptic sine function (see. e.g.,
Ref. \cite{AS}),
\begin{equation}\label{4-41}
\begin{split}
   &V=\tfrac12\sum_{i=1}^4\lambda_i\; ,\quad
    m=
\frac{(\lambda_2-\lambda_1)(\lambda_4-\lambda_3)}
{(\lambda_4-\lambda_2)(\lambda_3-\lambda_1)}\; ,
\end{split}
\end{equation}
and
\begin{equation}\label{4-42}
\begin{split}
C=& \tfrac18
(-\lambda_1-\lambda_2+\lambda_3+\lambda_4)\times \\
& (-\lambda_1+\lambda_2-\lambda_3+\lambda_4)\times
    (\lambda_1-\lambda_2-\lambda_3+\lambda_4) \; .\end{split}
\end{equation}
For constant $\lambda_i$'s, expressions \eqref{4-2},
\eqref{4-41} and \eqref{4-42} correspond to an exact (single phase)
solution of the NLS equation, periodic in time and space, where
oscillations have the amplitude
\begin{equation}\label{4-5}
    a=(\lambda_2-\lambda_1)(\lambda_4-\lambda_3)\; ,
\end{equation}
and the spatial wavelength
\begin{equation}\label{4-6}
L=
\frac{2\,\mathrm{K}(m)}{\sqrt{(\lambda_4-\lambda_2)(\lambda_3-\lambda_1)}} \; .
\end{equation}
In the limit $m\to 0$ ($\lambda_1=\lambda_2$ or
$\lambda_3=\lambda_4$), $\mathrm{sn}(x,m)\to \sin(x)$ Eq. (\ref{4-2})
describes a small amplitude sinusoidal wave oscillating around a
constant background. In the other limiting case $m\to 1$
($\lambda_2=\lambda_3$), $\mathrm{sn}(x,m)\to \tanh(x)$
and  Eq. (\ref{4-2}) describes a dark soliton (for which
$L\to\infty$).

The great insight of Gurevich and Pitaevskii \cite{gp-73} has been to
describe a dispersive shock wave as a slowly modulated nonlinear
wave, of type \eqref{4-2}, for which the $\lambda_i$'s are functions
of $x$ and $t$ which vary weakly over one wavelength and one
period. Their slow evolution is governed by the Whitham equations
\cite{whitham-74,kamch-2000}
\begin{equation}\label{4-7}
    \prt_t\lambda_i+
v_i(\lambda_1,\lambda_2,\lambda_3,\lambda_4)\, \prt_x\lambda_i=0,
\quad i=1,2,3,4.
\end{equation}
Comparing with Eqs.~(\ref{3-3c1}) one sees that the $\lambda_i$'s are
the Riemann invariants of the Whitham equations first found in
Refs.~\cite{fl86,pavlov87}.  The $v_i$'s are the associated
characteristic velocities; their explicit expressions can be obtained
from the relation \cite{Gur92,kamch-2000}
\begin{equation}\label{vi}
v_i= V-\frac 12 \frac{L}{\partial_i L} ,\quad
i=1,2,3,4\; ,
\end{equation}
where $\partial_i=\partial/\partial\la_i$. This yields
\begin{equation}\label{eq28}
    \begin{array}{l}
\displaystyle{
    v_1=V-\frac{(\la_4-\la_1)(\la_2-\la_1)\mathrm{K}(m)}
    {(\la_4-\la_1)\mathrm{K}(m)-(\la_4-\la_2)\mathrm{E}(m)},}\\[5mm]
\displaystyle{
    v_2=V+\frac{(\la_3-\la_2)(\la_2-\la_1)\mathrm{K}(m)}
    {(\la_3-\la_2)\mathrm{K}(m)-(\la_3-\la_1)\mathrm{E}(m)},}\\[5mm]
\displaystyle{
    v_3=V-\frac{(\la_4-\la_3)(\la_3-\la_2)\mathrm{K}(m)}
    {(\la_3-\la_2)\mathrm{K}(m)-(\la_4-\la_2)\mathrm{E}(m)},}\\[5mm]
\displaystyle{
    v_4=V+\frac{(\la_4-\la_3)(\la_4-\la_1)\mathrm{K}(m)}
    {(\la_4-\la_1)\mathrm{K}(m)-(\la_3-\la_1)\mathrm{E}(m)},}
    \end{array}
\end{equation}
where $m$ is given by Eq. \eqref{4-41} and $\mathrm{E}(m)$ is the complete
elliptic integrals of the second kind.

In the soliton limit $m \to 1$ (i.e., $\la_3 \to \la_2$), the Whitham
velocities reduce to
\begin{equation}\label{eq29}
   \begin{split}
&    v_1=\tfrac12(3\la_1+\la_4),\quad v_2=v_3=\tfrac12(\la_1+2\la_2+\la_4),\\
& v_4=\tfrac12(\la_1+3\la_4).
    \end{split}
\end{equation}
In a similar way, in the small amplitude limit $m\to0$ (i.e.,
$\la_2\to\la_1$), we obtain
\begin{equation}\label{eq30}
\begin{split}
& v_1=v_2=2\la_1+\frac{(\la_4-\la_3)^2}{2(\la_3+\la_4-2\la_1)},\\
& v_3=\tfrac12(3\la_3+\la_4),\quad v_4=\tfrac12(\la_3+3\la_4),
\end{split}
\end{equation}
and in another small amplitude limit ($m\to0$ when $\la_3\to\la_4$), we have
\begin{equation}\label{eq31}
\begin{split}
&    v_1=\tfrac12(3\la_1+\la_2),\quad v_2=\tfrac12(\la_1+3\la_2),\\
&    v_3=v_4=2\la_4+\frac{(\la_2-\la_1)^2}{2(\la_1+\la_2-2\la_4)}.
   \end{split}
\end{equation}

\subsection{Generalized hodograph method}\label{sec:GHM}

In Sec.~\ref{DSE} we have provided a nondispersive description of the
spreading and splitting of the initial pulse in two parts (one
propagating to the left, and the other to the right).  During this
nonlinear process the leading wavefront steepens and leads to wave
breaking. This occurs at a certain time $t_{\sss\rm WB}$ after which
the approach of Sec \ref{DSE} predicts a nonphysical multivalued
profile [see, e.g., Fig.~\ref{fig7c}], since it does not take into
account dispersive effects. The process of dispersive regularization
of the gradient catastrophe leads to the formation of a dispersive
shock wave, as first predicted by Sagdeev in the context of
collisionless plasma physics, see, e.g., Ref.~\cite{Moi63}.

For the specific case we are interested in, the Gurevich-Pitaevskii
approach which consists in using Whitham theory for describing the DSW
as a slowly modulated nonlinear wave holds, but it is complicated by
the fact that two of the four Riemann invariants vary in the shock
region. As already explained in the introduction, we adapt here the
method developed in Refs.~\cite{Gur91,Gur92,Kry92,ek-93} for treating
a similar situation for the Korteweg-de Vries equation.  The general
case of NLS dispersive shock with all four Riemann invariants varying
was considered in Ref.~\cite{ek-95}.

In all the following we concentrate our attention on the shock formed at
the right edge of the pulse propagating to the right. Due to the
symmetry of the problem the same treatment can be employed for the left
pulse. The prediction of multivalued
$\la^+$ resulting from the dispersionless approach of Sec.~\ref{DSE}
suggests that after wave breaking of the simple-wave solution,
the correct Whitham-Riemann invariant should be sought
in a configuration such that $\lambda_1=\la^-=-c_0$,
$\lambda_2=\la^+(x\to \infty)=c_0$ and $\lambda_3$ and $\lambda_4$
both depend on $x,t$. In this case the Whitham Eqs.~\eqref{4-7} with
$i=1,\,2$ are trivially satisfied, and for solving them for $i=3$ and
4, one introduces two functions $W_i(\la_3,\la_4)$ ($i=3$ or 4),
exactly as we did in Sec.~\ref{DSE} with $W_{\pm}(\la^-, \la^+)$:
\begin{equation}\label{ek17}
x-v_i(\la_3,\la_4) t =W_i(\la_3,\la_4), \quad i=3,4\; .
\end{equation}
For the sake of brevity we have denoted in the above equation
$v_i(\la_3,\la_4)=v_i(\la_1=-c_0,\la_2=c_0,\la_3,\la_4)$ for
$i\in\{3,4\}$; we will keep this notation henceforth.

\begin{figure}
\centering
\includegraphics[width=\linewidth]{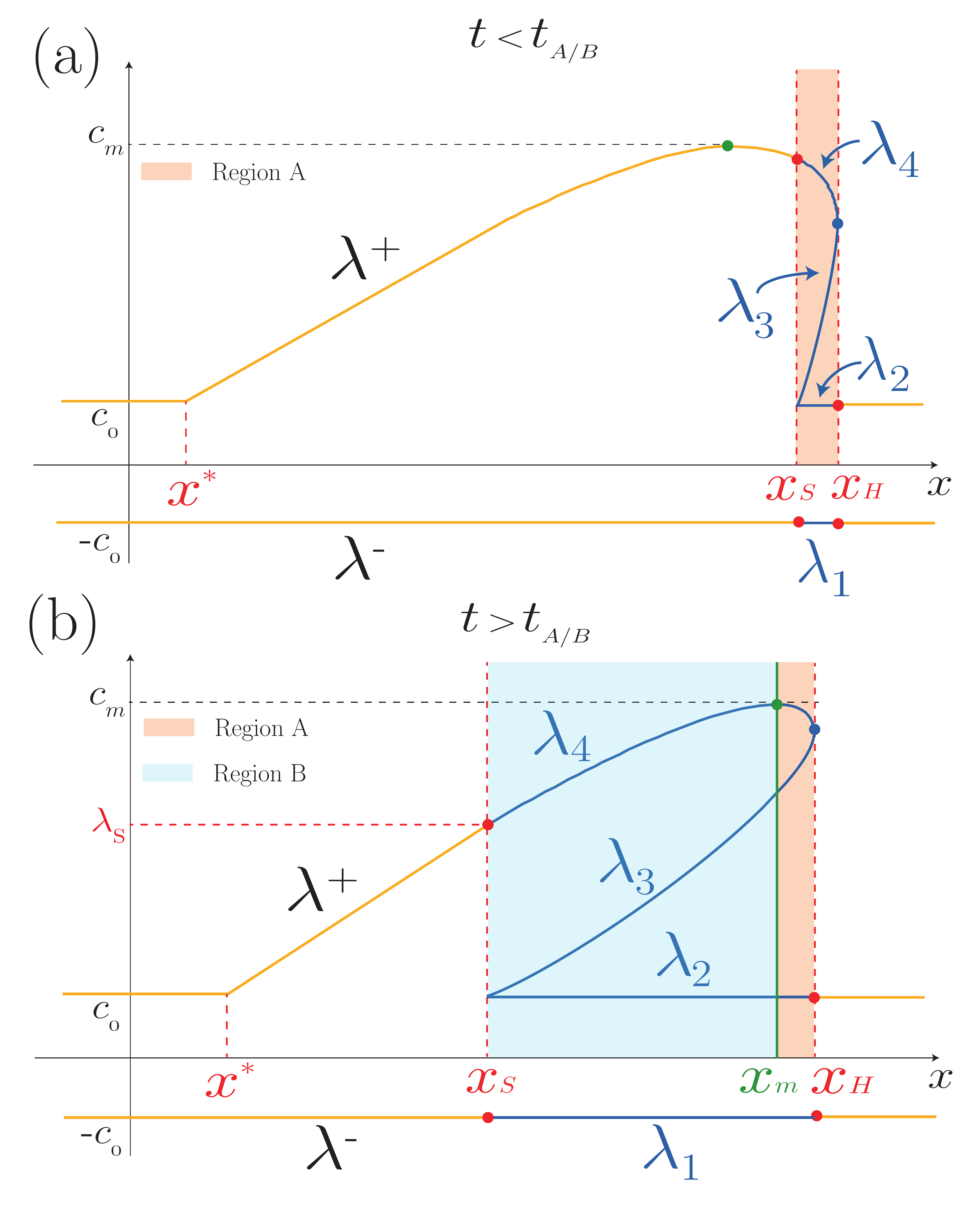}
\caption{Schematic plots of the position dependence of the Riemann
  invariants inside (blue solid curves) and outside (yellow solid
  curves) the DSW (colored region).  (a) For $t<t_{\sss\rm A/B}$, the
  DSW is connected to the smooth profile coming from the time
  evolution of part A of the initial pulse. At $t=t_{\sss\rm A/B}$,
  part A is completely absorbed by the DSW. Thus, for this time, the
  shock wave connects to the smooth profile exactly at
  $\la_+(\xsol(t),t)=c_m$. (b) For $t>t_{\sss\rm A/B}$, the DSW is
  connected at its left edge at a point belonging to part B of the
  dispersionless profile. In this case the shock wave is divided in
  two regions A and B, separated by the green vertical line in the
  plot. The continuity along the separation line between the two
  regions (i.e., at $x=x_m(t)$) is ensured by Eq.~\eqref{ek21}.}
\label{fig8}
\end{figure}

Then, one can derive Tsarev equations for $W_i(\la_3,\la_4)$
[replacing the subscripts $+$ and $-$ by $4$ and $3$ in
\eqref{Tsarev}] and one can show (see, e.g.,
Refs.~\cite{Gur92,ek-95,wright,tian}) that these are solved for $W_i$'s of
the form
\begin{equation}\label{ek19}
W_i = \left(1-\frac{L}{\prt_{i}L}\prt_{i}\right)\mathscr{W}
=\mathscr{W}+ 2\,(v_i - V) \prt_{i} \mathscr{W}\; ,
\end{equation}
where $\mathscr{W}(\la_3,\la_4)$ is solution of the Euler-Poisson equation
\begin{equation}\label{ek20}
\partial_{34}\mathscr{W}=
\frac{\prt_3\mathscr{W}-\prt_4\mathscr{W}}{2(\la_3-\la_4)}\; .
\end{equation}
As was first understood in Ref.~\cite{gkm-89}, after the wave breaking
time, the development of the dispersive shock wave occurs in {\it two}
steps. Initially (when $t$ is close to
$t_{\sss\rm WB}$), the DSW is connected at its left edge to the
smooth profile coming from the time evolution of the right part of
the initial profile of $\la^+$ (part A), which is gradually absorbed
in the DSW. This process of absorption is complete at a time we
denote as $t_{\sss\rm A/B}$. Then, for $t>t_{\sss\rm A/B}$, the DSW
is connected to the smooth profile coming from the time evolution of
part B of $\la^+$ (this is case ``B'', ``region B'' of the $(x,t)$
plane). During the initial step (for $t<t_{\sss\rm A/B}$), for a
given time $t$, the highest value of the largest of the Riemann
invariant is reached within the smooth part of the profile and keeps
the constant value $c_m$. Then, in the subsequent time evolution,
this highest value is reached within the DSW (or at its right
boundary) where there exists a point where $\la_4$ takes its maximal
value ($c_m$). We illustrate these two steps of development of the
DSW in Fig.~\ref{fig8}. We denote the region of the DSW where $\la_4$
is a decreasing function of $x$ as region A, the part where it
increases as region B.

In region A of the $(x,t)$ plane, we denote by
$\mathscr{W}^{\sss\rm A}(\la_3,\la_4)$ the solution of the Euler-Poisson
equation, in region B we denote it instead as
$\mathscr{W}^{\sss\rm B}(\la_3,\la_4)$. These two forms are joined by the line
$\la_4=c_m$ where
\begin{equation}\label{ek21}
\mathscr{W}^{\sss\rm A}(\la_3,c_m)=\mathscr{W}^{\sss\rm B}(\la_3,c_m)\; .
\end{equation}
We denote the position where this matching condition is realized as
$x_{m}(t)$, see Fig.~\ref{fig8}(b). The corresponding boundary in the
$(x,t)$ plane is represented as a green solid line in Fig.~\ref{fig9}.

Since the general solution of the Euler-Poisson equation with the
appropriate boundary conditions and the construction of the resulting
nonlinear pattern are quite involved, we shall first consider some
particular---but useful---results which follow from general principles of
the Whitham theory.

\section{Motion of the soliton edge of the shock}\label{sol-edge}

During the first stage of evolution of the DSW, its left (solitonic)
edge is connected to the smooth dispersionless solution whose dynamics
is described by formula (\ref{IIIc}), that is, we have here
\begin{equation}\label{eq1ak}
  \xsol - v_{+}(-c_0,\lsol)\,t=W_{+}^{(3)}(-c_0,\lsol),
\end{equation}
where $\xsol(t)$ is the position of the left edge of the DSW and
$\lsol(t)\equiv\la^+(\xsol(t),t)$. We recall that in all the following we
focus on the DSW formed in the right part of the pulse. Hence the
above equation concerns the right part of the nondispersive part of
the profile. According to the terminology of section \ref{DSE}, this
corresponds to region III.

On the other hand, in vicinity of this boundary, the Whitham equations
(\ref{4-7}) with the limiting expressions (\ref{eq29}) (where
$\la_2=\la_3=c_0$) for the velocities $v_i$ are given by
\begin{equation}\label{t3-148.1}
\begin{split}
 & {\prt_t}\la_3+\frac{1}{2} \left( \la_4 + c_0\right){\prt_x}\la_3=0,\\
 & {\prt_t}\la_4+\frac{1}{2}\left( 3\la_4 - c_0\right){\prt_x}\la_4=0.
   \end{split}
\end{equation}
For solving these equations one can perform a classical hodograph
transform, that is, one assumes that $x$ and $t$ are functions of
the independent variables $\la_3$ and $\la_4$: $t=t(\la_3,\la_4)$ and
$x=x(\la_3,\la_4)$. We find from Eqs.~\eqref{t3-148.1} that these
functions must satisfy the linear system
\begin{equation}\nonumber
\begin{split}
&    \frac{\prt x}{\prt \la_3}-\frac{1}{2}
\left( 3\la_4 - c_0\right)\frac{\prt t}{\prt \la_3}=0,\\
&    \frac{\prt x}{\prt \la_4}-\frac{1}{2}
\left( \la_4 + c_0\right)\frac{\prt t}{\prt \la_4}=0.
   \end{split}
\end{equation}
At the left edge of the DSW, the second equation reads
\begin{equation}\label{t3-148.3}
  \frac{\prt \xsol}{\prt \lsol}
-\frac{1}{2} \left( \lsol + c_0\right)\frac{\prt t}{\prt \lsol}=0,
\end{equation}
and this must be compatible with Eq.~(\ref{eq1ak}). Differentiation of
Eq.~(\ref{eq1ak}) with respect to $\lsol$ and elimination of
$\prt \xsol/\prt \lsol$ with the use of
Eq.~\eqref{t3-148.3} yields a differential equation for the function
$t(\lsol)\equiv t(c_0,\lsol)$:
\begin{equation}\label{t3-148.5}
 \left( \lsol - c_0 \right)\frac{dt}{d \lsol}+\frac{3}{2}\,t=
-\frac{dW_{+}^{(3)}(-c_0,\lsol)}{d \lsol}.
\end{equation}
At the wave breaking time, $\lsol=c_0$, this corresponds to the
definition $t_{\rm\sss WB}=t(c_0)$ and Eq.~\eqref{t3-148.5} then
yields
\begin{equation}\label{eq:tWB}
t_{\rm\sss WB}=-\frac{2}{3}
\left.\frac{dW_{+}^{(3)}(-c_0,\lsol)}{d \lsol}\right|_{\lsol=c_0}\; ,
\end{equation}
in agreement with Eq.~(\ref{tWB-est}), what should be expected since
at the wave breaking moment the DSW reduces to a point in the Whitham
approximation. For the concrete case of our initial distribution we
can get a simple explicit expression for $t_{\sss\rm WB}$ which reads
(see Eq.~\eqref{tWB-est} and note \cite{foot1}):
\begin{equation}\label{eq:tWB_approx}
t_{\rm\sss WB}= -\frac{2}{3}
\left.\frac{dw^{\sss\rm A}(\lsol)}{d \lsol}\right|_{\lsol=c_0}
=\frac{2\, c_0\,x_0}{3\, \rho_1}\; ,
\end{equation}
where the right hand side is the form of
the central formula corresponding to the initial profile
\eqref{rho_init}. Taking $\rho_0 = 0.5$, $\rho_m=2$ and $x_0=20$, we
find $t_{\rm\sss WB} \simeq 6.3$, in excellent agreement with the
numerical simulations.

The solution of Eq.~\eqref{t3-148.5} reads
\begin{equation}\label{t3-148.6}
\begin{split}
  t(\lsol)& =\frac{-1}{(\lsol-c_0)^{3/2}}
\int_{c_0}^{\lsol}\!\!\!\sqrt{r-c_0}\,\frac{dW_{+}^{(3)}(-c_0,r)}{d r}\, dr\\
& = \frac1{2(\lsol-c_0)^{3/2}}
\int_{c_0}^{\lsol} \frac{W_{+}^{(3)}(-c_0,r)}{\sqrt{r-c_0}}dr\\
&
-\frac{W_{+}^{(3)}(-c_0,\lsol)}{\lsol-c_0}.
\end{split}
\end{equation}

Substituting this expression into (\ref{eq1ak}), we
obtain the function $\xsol(\lsol)\equiv x(c_0,\lsol)$:
\begin{equation}\label{t3-148.7}
\xsol(\lsol) = \frac{1}{2}\left( 3\lsol-c_0\right)\,t(\lsol)
+ W_{+}^{(3)}(-c_0,\lsol).
\end{equation}
The two formulas \eqref{t3-148.6} and \eqref{t3-148.7} define in an
implicit way the law of motion $x=\xsol(t)$ of the soliton
edge of the DSW.

The above expressions are correct as long as the soliton edge is
located inside region A of the DSW, that is, up to the moment
$t_{\sss\rm A/B}=t(c_m)$. From \eqref{t3-148.6} one obtains the
explicit expression
\begin{equation}\label{t3-148.6ak}
  t_{\sss\rm A/B}
  =\frac{-1}{(c_m-c_0)^{3/2}}
\int_{c_0}^{c_m}\!\!\!\sqrt{r-c_0}\,\frac{dW_{+}^{(3)}(-c_0,r)}{d r}\, dr.
\end{equation}
In the case we consider this yields $t_{\sss\rm A/B}=25.9$.  For time
larger than $t_{\sss\rm A/B}$ the soliton edge connects with region B
of the dispersionless profile which corresponds to region II$_r$, see
Fig.~\ref{fig.char-traj}.  Concretely, for a time $t>t_{\sss\rm A/B}$,
instead of Eq. \eqref{t3-148.5} we have to solve the differential
equation
\begin{equation}\label{toto}
 \left( \lsol - c_0 \right)\frac{dt}{d \lsol}+\frac{3}{2}\,t=
-\frac{dW_{+}^{(2)}(-c_0,\lsol)}{d \lsol}.
\end{equation}
with the initial condition $t(c_m)=t_{\sss\rm A/B}$.  The solution of
Eq.~\eqref{toto} reads
\begin{equation}\label{t3-148.8a}
\begin{split}
  t(\lsol)=&\frac{-1}{(\lsol-c_0)^{3/2}}
\Bigg(\int_{c_m}^{\lsol}\!\!\sqrt{r-c_0}\,\frac{dW_{+}^{(2)}(-c_0,r)}{d r}dr\\
  &+\int_{c_0}^{c_m}\sqrt{r-c_0}\,\frac{dW_{+}^{(3)}(-c_0,r)}{d r}dr\Bigg),
\end{split}
\end{equation}
and $\xsol(\lsol)$ is determined by Eq. \eqref{IIIe}:
\begin{equation}\label{t3-148.8b}
\xsol(\lsol) = \frac{1}{2}\left( 3\lsol-c_0\right)\,t(\lsol)
+ W_{+}^{(2)}(-c_0,\lsol).
\end{equation}
At asymptotically large time $t\to\infty$ one is in stage B of
evolution of the DSW with furthermore $\lsol\to c_0$.  In this case
the upper limit of integration in the first integral of formula
\eqref{t3-148.8a} can be put equal to $c_0$. Thus, we get in this
limit
\begin{equation}
t(\lsol)\simeq \frac{\cal A}{(\lsol -c_0)^{3/2}},
\end{equation}
where the expression for the constant ${\cal A}$ is
\begin{equation}\label{bigA}
\begin{split}
 \cal A=&
- \Bigg(\int_{c_m}^{c_0}\sqrt{r-c_0}\,\frac{dW_{+}^{(2)}(-c_0,r)}{d r}dr\\
  &+\int_{c_0}^{c_m}\sqrt{r-c_0}\,\frac{dW_{+}^{(3)}(-c_0,r)}{d r}dr\Bigg).
\end{split}
\end{equation}
Consequently one obtains the asymptotic expressions
\begin{equation}\label{xrrr}
\lsol(t)=c_0+\left(\frac{\cal A}{ t}\right)^{2/3}\; ,\quad
 \xsol(t)=c_0\,t + \frac{3{\cal A}^{2/3}}{2}\, t^{1/3}\; .
\end{equation}
We denote the position of the rear point of the simple wave as
$x^*(t)$, see Fig.~\ref{fig8}. It is clear from
Fig.~\ref{fig.char-traj} that $x^*=0$ at time
$t=t _{\sss\rm SW}(c_0)$, i.e., just when region 2 disappears, whereafter
the dispersionless approach of Sec.~\ref{DSE} predicts a profile
with only simple waves and plateau regions. The rear edge of the simple
wave then propagates over a flat background at constant velocity
$c_0$; one thus has
\begin{equation}
x^*(t)=c_0 \left[t-t_{\sss\rm SW}(c_0)\right]\; .
\end{equation}
Asymptotically (i.e., at time much larger than $t_{\sss\rm SW}(c_0)$)
one has $x^*(t)\simeq c_0 t$ and, in the simple wave profile between
$x^*(t)$ and $\xsol(t)$, $\lambda^+$ depends on the self-similar
variable $(x-x^*(t))/t$ while $\lambda^-$ is constant. Then
Eqs.~\eqref{3-3c1} readily yield
\begin{equation}
\begin{cases}
\lambda^+=c_0+\frac23 \frac{\displaystyle x-x^*(t)}{\displaystyle t},\\
\lambda^-=-c_0,
\end{cases}
\mbox{for}\; x\in [x^*(t),\xsol(t)]\; .
\end{equation}
Eqs.~\eqref{3-3b} then yield the explicit expression of $\rho$ in this
region (which was roughly described in the end of Sec.~\ref{sec:model}
as having a ``quasi-triangular shape''), and using \eqref{xrrr} one
obtains
\begin{equation}\label{conserved}
\int_{x^*(t)}^{\xsol(t)}\left(\sqrt{\rho(x,t)}-c_0\right)^{1/2}dx
=\frac{1}{\sqrt{2}} \, {\cal A} \; .
\end{equation}
The asymptotic situation at the rear of the DSW is reminiscent of what
occurs in the theory of weak dissipative shocks where (i) a nonlinear
pattern of triangular shape may also appear at the rear edge of a
(viscous) shock, (ii) the details of the initial distribution are lost
at large time (as in the present case) and (iii) a conserved quantity
of the type \eqref{conserved} also exists. Hence the above results
provide, for a conservative system, the counterpart of the weak
viscous shock theory (presented for instance in
Ref.~\cite{whitham-74}). Note, however, that the boundary conditions
at the large amplitude edge of the shock are different depending on
whether one considers a dissipative or a conservative system, and that
the corresponding velocity and conserved quantity are accordingly also
different. Note also that equivalent relations for the behavior of a
rarefaction wave in the rear of a dispersive shock in the similar
situation for the Korteweg-de Vries equation has been obtained in
Ref.~\cite{Iso18}.

Formulas \eqref{xrrr} and \eqref{conserved} are important because they
provide an indirect evidence making it possible to assert if a given
experiment has indeed reached to the point where a {\it bona fide}
dispersive shock wave should be expected.

We now give, in the next section, the explicit theoretical description
of the whole region of the dispersive shock.

\section{Solution in the shock region}\label{full}

In this section we turn to the general solution of the Whitham
equations given by the formulas of Sec.~\ref{sec:GHM}. Our task is
to express the functions $W_3$ and $W_4$ in terms of the initial
distribution of the light pulse. As was indicated above, one needs to
distinguish two regions, A and B, in which $\mathscr{W}$ takes
different values.

\subsection{Solution in region A}

In region A one can straightforwardly adapt the procedure explained in
Ref.~\cite{Gur92}. One imposes the matching of the left edge of the
DSW with the dispersionless solution (see Sec.~\ref{sec.match}): just
at $x=\xsol(t)$, we have $\la_4=\la^+$, $\la_3=\la_2 = c_0$,
$\la_1 =-c_0$ (see Fig. \ref{fig8}) and Eq.~\eqref{eq29} yields
$v_4(\la_3,\la_4)=(3\,\la_4-c_0)/2=v_+(-c_0,\lambda^+)$. Then, at this
point, the conditions \eqref{IIIc} and \eqref{ek17} with $i=4$ are
simultaneously satisfied which implies
\begin{equation}\label{ek22}
W^{\sss\rm A}_4(\la_3 = c_0,\la_4=
\la^+) =W_{+}^{(3)}(-c_0,\la^+)\; ,
\end{equation}
where $W_{+}^{(3)}$ is the form of $W_+$ corresponding to region~3.
Note that, here, the first argument of the function $W_{+}^{(3)}$
is $\lambda^- = -c_0$ for all time. Indeed, the boundary condition
\eqref{ek22} corresponds to the matching in physical space at
$\xsol(t)$. When the DSW starts to form at time $t_{\sss\rm WB}$,
the edge $\xsol(t_{\sss\rm WB})$ lies on the characteristic issued from $x_0$
[$x_0$ defines the initial extend of the pulse,
see. Eq.~\eqref{rho_init}].  The Riemann invariant $\la^-$ is constant
and equal to $-c_0$ along this characteristic,
cf. Fig.~\ref{fig.char-traj}. Then, because the characteristics of
$\la^-$ in the dispersionless region close to $\xsol$ are oriented to
the left whereas $\xsol$ moves to the right, it is clear that
$\la^-(\xsol(t),t) = -c_0$ for $t \geq t_{\sss\rm WB}$.

In terms of
$\mathscr{W}$ the relation \eqref{ek22} corresponds to the equation
\begin{equation}\label{ek23}
\mathscr{W}^{\sss\rm A} (c_0, \la_4)  + 2\,(\la_4 - c_0)
\partial_{4} \mathscr{W}^{\sss\rm A}(c_0, \la_4) =W_{+}^{(3)}(-c_0,\la_4)\; ,
\end{equation}
whose solution is
\begin{equation}\label{ek24}
\mathscr{W}^{\sss\rm A} \left(c_0, \lambda_4\right) =
\frac{1}{2\, \sqrt{\lambda_4 - c_0}}
\int_{c_0}^{\lambda_4} \frac{W^{(3)}_+(-c_0,r)\, dr}{\sqrt{r - c_0}}\; .
\end{equation}
This will serve as a boundary condition for the
Euler-Poisson equation \eqref{ek20} whose general solution has
been given by Eisenhart \cite{Eis18} in the form
\begin{equation}\label{ek25}
\begin{split}
\mathscr{W}^{\sss\rm A}(\lambda_3, \lambda_4) =
&\int_{c_0}^{\lambda_3}
\frac{\psi^{\sss\rm A}(\mu)\, d\mu}{\sqrt{\lambda_3-\mu}\sqrt{\lambda_4-\mu}}
+\\
& \int_{c_0}^{\lambda_4}
\frac{\varphi^{\sss\rm A}(\mu)\, d\mu}
{\sqrt{|\lambda_3-\mu|}\sqrt{\lambda_4-\mu}} \; ,
\end{split}
\end{equation}
where $\varphi^{\sss\rm A}(\mu)$ and $\psi^{\sss\rm A}(\mu)$ are
arbitrary functions to be determined from the appropriate boundary
conditions. By taking $\la_3 = c_0$ in this expression one sees that
$\varphi^{\sss\rm A}(\mu)/\sqrt{\mu-c_0}$ is the Abel transform of
$\mathscr{W}^{\sss\rm A}(c_0,\la_4)$. Using the inverse
transformation \cite{Ark05} and expression \eqref{ek24} one can show that
\begin{equation}\label{ek27}
\varphi^{\sss\rm A}(\mu) = \frac{1}{2\pi\,\sqrt{\mu - c_0}}
\int_{c_0}^{\mu} \frac{W^{(3)}_+(-c_0,r) \, dr}{\sqrt{\mu - r }} \; ,
\end{equation}
where we recall that $W^{(3)}_+=\partial_+\chi^{(3)}$.  In order to
determine the other unknown function $\psi^{\sss\rm A}$, one considers
the right boundary of the DSW where $\la_3$ and $\la_4$ are
asymptotically close to each other. One can show (see, e.g., an
equivalent reasoning in Ref.~\cite{Iso18}) that in order to avoid
divergence of $\mathscr{W}^{\sss\rm A}(\lambda_3,\la_4=\la_3)$, one
needs to impose
$\psi^{\sss\rm A}(\lambda) = -\varphi^{\sss\rm A}(\lambda)$. The final
form of the Eisenhart solution in region A thus reads
\begin{equation}\label{ek28}
\mathscr{W}^{\sss\rm A}(\lambda_3, \lambda_4) =
\int_{\lambda_3}^{\lambda_4}
\frac{\varphi^{\sss\rm A}(\mu) \, d\mu}
{\sqrt{\mu - \lambda_3}\sqrt{\lambda_4 - \mu}}\; ,
\end{equation}
where $\varphi^{\sss\rm A}$ is given by formula~\eqref{ek27}.

\subsection{Solution in region B}

One looks for a solution of the Euler-Poisson equation in
region B in the form
\begin{equation}\label{ek29}
\mathscr{W}^{\sss\rm B}(\lambda_3, \lambda_4) =
\mathscr{W}^{\sss\rm A}(\lambda_3, \lambda_4) +
\int_{\lambda_4}^{c_m} \!\!\!
\frac{\varphi^{\sss\rm B}(\mu) \, d\mu}{
\sqrt{\mu - \lambda_3}\sqrt{\mu - \lambda_4}},
\end{equation}
where $c_m$, the maximum value for $\la_4$.  The above expression
ensures that $\mathscr{W}^{\sss\rm B}$, (i) being the sum of two
solutions of the Euler-Poisson equation, is also a solution of this
equation and (ii) verifies the boundary condition \eqref{ek21} since
the second term of the right-hand side of \eqref{ek29} vanishes when
$\la_4=c_m$.

At the left boundary of the DSW, $\mathscr{W}^{\sss\rm B}(c_0,\la_4)$
verifies an equation similar to \eqref{ek23}:
\begin{equation}
\mathscr{W}^{\sss\rm B} (c_0, \la_4)  + 2\,(\la_4 - c_0)
\partial_{4} \mathscr{W}^{\sss\rm B}(c_0, \la_4) =W_{+}^{(2)}(-c_0,\la_4)\; ,
\end{equation}
The
solution with the appropriate integration constant reads
\begin{equation}\label{ek30}
\begin{split}
\mathscr{W}^{\sss\rm B}(r_1,0) & = \frac{1}{2\, \sqrt{\lambda_4 - c_0}}
\int_{\la_4}^{c_m}
\frac{W^{(2)}_+(-c_0,r)}{\sqrt{r - c_0}} dr\\
&+\frac{1}{2\, \sqrt{\lambda_4 - c_0}} \int_{c_0}^{c_m}
\frac{W^{(3)}_+(-c_0,r)}{\sqrt{r - c_0}} dr\; ,
\end{split}
\end{equation}
where $W^{(2)}_+$ is the form of $W_+$ corresponding to region
2. The same procedure as the one previously used for the part A of the DSW
leads here to
\begin{equation}\label{ek31}
\varphi^{\sss\rm B}(\mu) = \frac{1}{2\pi\,\sqrt{\mu - c_0}}
\int_{\mu}^{c_m} \frac{W^{(3)}_+(-c_0,r) - W^{(2)}_+(-c_0,r)}
{\sqrt{\mu - r }} dr \; .
\end{equation}
Eqs.~\eqref{ek29} and \eqref{ek31} give the
solution of the Euler-Poisson equation in region B.

\subsection{Characteristics of the DSW at its
edges}\label{sec:Boundaries}

It is important to determine the boundaries $\xsol(t)$ and
$x_{\sss\rm H}(t)$ of the DSW, as well as the values of the
Riemann invariants $\la_3$ and $\la_4$ at these points. The law of motion
of the soliton edge was already found in Sec.~\ref{sol-edge} and
it is instructive to show how this result can be obtained from the
general solution.

At the soliton edge we have $\la_2=\la_3=c_0$ and $\la_4=\lsol(t)$.
The corresponding Whitham velocities are $v_3=(\lsol+c_0)/2$ and
$v_4=(3\lsol-c_0)/2$ [see Eqs.~(\ref{eq29})], and the two
equations \eqref{ek17} read
\begin{equation}\label{ek32}
\begin{split}
\xsol-\frac{1}{2}\,(3\lsol-c_0) t =W_4^\alpha(c_0,\lsol)&
= W_+^{(n)}(-c_0,\lsol), \\
\xsol-\frac{1}{2}\,(\lsol+c_0) t =W_3^\alpha(c_0,\lsol)&
=\mathscr{W}^{\alpha}(c_0,\lsol),
\end{split}
\end{equation}
where, in order to have formulas applying to both stages of evolution
of the DSW, one has introduced dummy indices $\alpha$ and $n$ with
$\alpha=$ A or B and $n=3$ or 2, respectively. This gives
at once
\begin{equation}\label{35.28}
  \begin{split}
  & t(\lsol)=\frac1{\lsol-c_0}
\left[\mathscr{W}^\alpha(c_0,\lsol)-W_+^{(n)}(-c_0,\lsol)\right],\\
  & \xsol(\lsol)= c_0\,t
+ \frac12\left[3\mathscr{W}^\alpha(c_0,\lsol)-W_+^{(n)}(-c_0,\lsol)\right].
  \end{split}
\end{equation}
Let us consider the stage A for instance. Eq.~\eqref{ek24} yields
\begin{equation*}
\mathscr{W}^{A} \left(c_0, \lsol\right) =
\frac{1}{2\, \sqrt{\lsol - c_0}}
\int_{c_0}^{\lsol} \frac{W^{(3)}_+(-c_0,r)\, dr}{\sqrt{r - c_0}},
\end{equation*}
which, inserted into
Eqs.~(\ref{35.28}) gives immediately the results (\ref{t3-148.6}) and
(\ref{t3-148.7}).

\begin{figure}
\centering
\includegraphics[width=\linewidth]{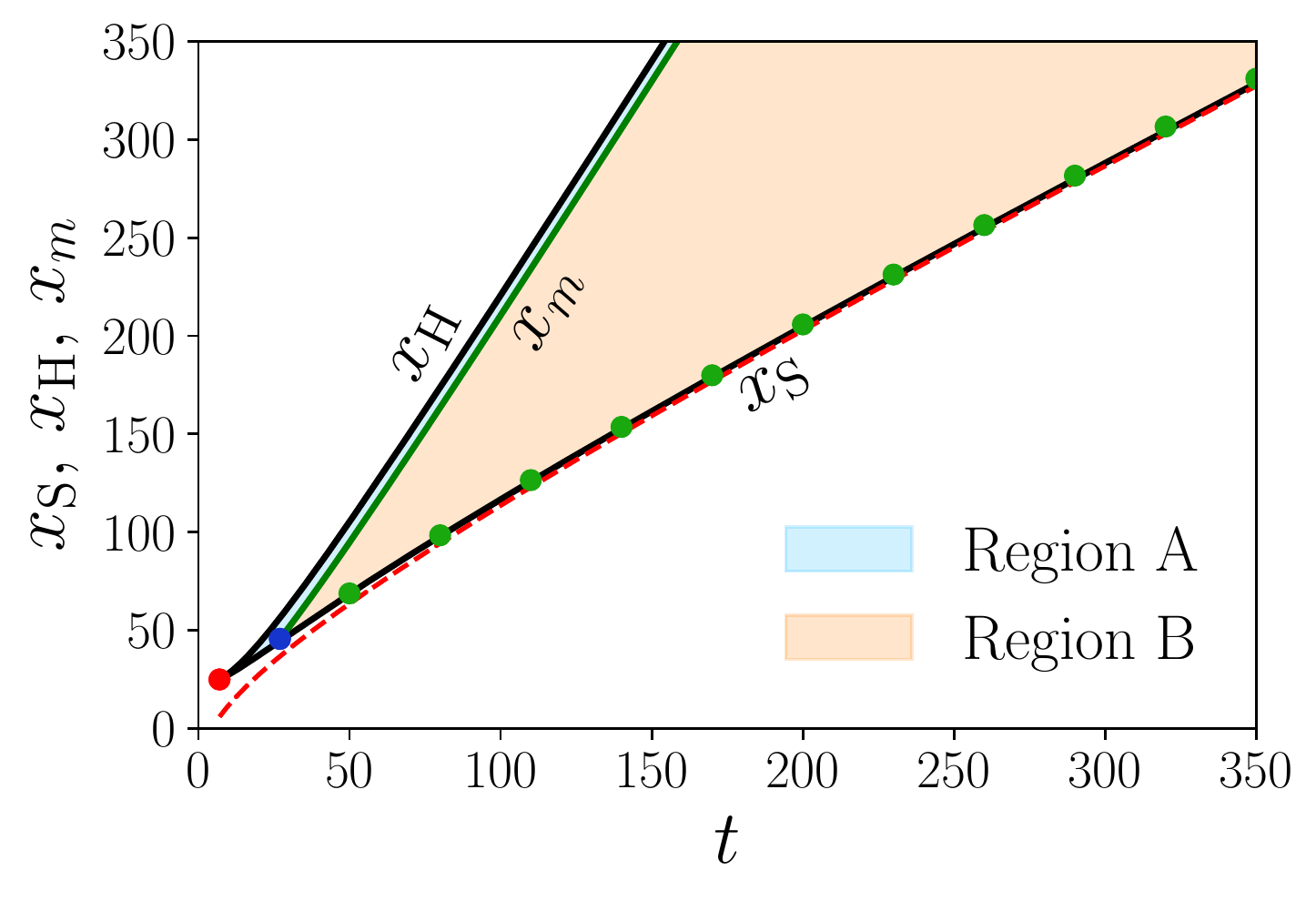}
\caption{ Black solid curves: time evolution of $\xsol(t)$ and
  $x_{\sss \rm H}(t)$ calculated from Eqs.~\eqref{35.28} and
  \eqref{ek34}. Green solid curve: time evolution of $x_{m}(t)$, for
  which $\la_4(x_{m}(t),t) = c_{m}$, which marks the separation
  between regions A and B. Red dashed line: asymptotic behavior of
  $\xsol(t)$, from Eq.~\eqref{xrrr}. The green points indicate the
  position $\xsol(t)$ extracted from simulations, for an initial
  condition \eqref{rho_init} with $\rho_0 = 0.5$, $\rho_m = 2$ and
  $x_0=20$. The red dot marks the birth of the DSW (at time
  $t_{\sss \rm WB}\simeq 6.3$), while the blue one initiates region B (at time
  $t_{\sss \rm A/B}\simeq 25.9$).}
\label{fig9}
\end{figure}

\begin{figure}
\centering
\includegraphics[width=\linewidth]{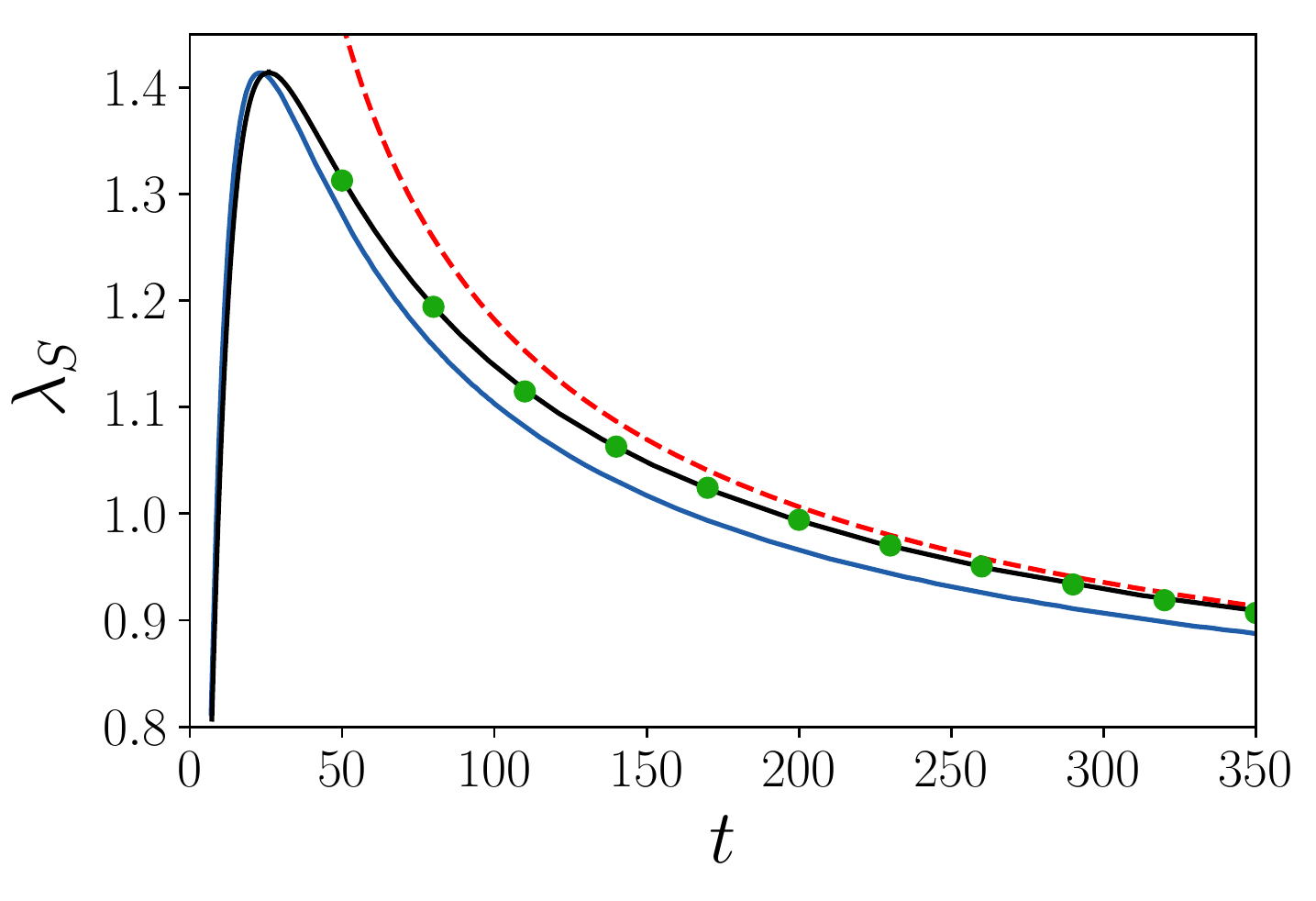}
\caption{Black solid curve: time evolution of $\lsol(t)$ from
  Eq.~\eqref{35.28}, or equivalently from Eqs. \eqref{t3-148.6} and
  then \eqref{t3-148.8a}. Red dashed line: asymptotic behavior, from
  Eq.~\eqref{xrrr}. The green points are extracted from simulations
  for different times, for the initial profile \eqref{rho_init} with
  $\rho_0 = 0.5$, $\rho_m = 2$ and $x_0=20$. The blue solid curve is
  an approximation obtained by schematically describing the initial
  splitting by assuming $\la^- \simeq \text{cst}$ for all $t$'s during
  the evolution of the right pulse (see the text).}
\label{fig10}
\end{figure}

Fig. \ref{fig9} shows the time evolution of $\xsol(t)$. The black
curve is calculated from Eqs.~(\ref{35.28}), while the red dashed line
corresponds to the asymptotic behavior of $\xsol$, given by
Eq.~\eqref{xrrr}. The green points are extracted from simulations and
exhibit a very good agreement with the theory.
The same excellent agreement is obtained for the time evolution of
$\lsol$, as shown in Fig.~\ref{fig10}.

We demonstrated in Sec.~\ref{DSE} the accuracy of the Riemann method
for describing the spreading and splitting of the initial pulse into
two parts. The matching between the left edge of the DSW and the
dispersionless profile at the point of coordinates
$\left(\xsol,\lsol \right)$ is given in Eq.~\eqref{eq1ak}. Since the
splitting occurs rapidly, a simpler approach would be to make the
approximation $\lambda^-(x,t) = -c_0 = \text{const}$ for the
dispersionless right part of the profile. In this case, the Riemann
equation \eqref{3-3c1} for $\lambda^+$ reduces to
\begin{equation}
\frac{\partial \lambda^+}{\partial t} + \left( \frac{3}{2} \la^+
- \frac{1}{2} c_0 \right) \frac{\partial \lambda^+}{\partial x} = 0.
\end{equation}
This equation can be solved by the method of characteristics which
yields the implicit solution for $\la^+(x,t)$:
\begin{equation}
x -  \left( \frac{3}{2} \la^+ - \frac{1}{2} c_0 \right) \, t
= w^{\sss\rm A/B}(\la^+)\; ,
\end{equation}
where $w^{\sss\rm A/B}$ are the the inverse functions of the initial
$\la^+(x)$ profile in parts A and B [in our case their explicit
expressions are given in Eqs.~\eqref{InvFunc}].

Within this approximation the DSW is described through $\mathscr{W}$
by the same equations \eqref{ek27}, \eqref{ek28}, \eqref{ek29} and
\eqref{ek31} as before, replacing $W_{+}^{(3/2)}(-c_0,r)$ by
$w^{\sss\rm A/B}(r)$ everywhere. $\lsol$ computed using this
approximation is represented in Fig.~\ref{fig10} as a function of $t$
(blue solid line) where it is also compared with the results obtained
using the full Riemann method (black solid line) and the results
extracted from numerical simulations (green dots). As we can see, an
accurate description of the spreading and splitting stage is important
since the blue curve does not precisely agree with the results of the
simulations, mainly at large times. However, this approximation gives
a correct description of the initial formation of the DSW: this is
discussed in note \cite{foot1} where it is argued that, close to the
wave breaking time, the approximation
$W_{+}^{(3)}(-c_0,r)\simeq w^{\sss\rm A}(r)$ is very accurate.

Let us now turn to the determination of the location
$x_{\sss\rm H}(t)$ of the small-amplitude, harmonic boundary of the DSW,
and of the common value $\la_{\sss\rm H}(t)$ of $\la_3$ and $\la_4$ at
this point (see Fig.~\ref{fig8}). In the typical situation the left
boundary is located in region A. In this case the equations
\eqref{ek17} for $i=3$ and 4 are equivalent and read
\begin{equation}\label{ek34}
x_{\sss\rm H}-v_{\sss\rm H} \cdot t =W_i^{\sss\rm A}(\la_{\sss\rm H},\la_{\sss\rm H})\; ,
\quad i=3\;\mbox{or}\;4\; ,
\end{equation}
where
$v_{\sss\rm H} = v_{i}( \la_{\sss\rm H}, \la_{\sss\rm H}) = 2\,
\la_{\sss\rm H} - c_0^2/\la_{\sss\rm H}$
[cf. Eqs.~\eqref{eq31}]. An equation for $\la_{\sss\rm H}$ alone is
obtained by demanding that the velocity $d x_{\sss\rm H}/d t$ of the
left boundary is equal to the common value $v_{\sss\rm H}$ of $v_3$
and $v_4$. The differentiation of Eq.~\eqref{ek34} with respect to
time then yields
\begin{equation}\label{ek35}
t =-\frac{1}{{d}v_{\sss\rm H}/ {d} \la_{\sss\rm H}  }\,
\frac{{d}W_4^{\sss\rm A}(\la_{\sss\rm H},\la_{\sss\rm H})}{{d}\la_{\sss\rm H}}\; .
\end{equation}

Note that the relation $d x_{\sss\rm H}/{d}t=v_{\sss\rm H}$
is a consequence of the general statement that the small amplitude
edge of the DSW propagates with the group velocity corresponding to
the wave number determined by the solution of the Whitham
equations. Indeed, the NLS group velocity of a linear wave with
wave-vector $k$ moving over a background $\rho_0 = c_0^2$ is the
group velocity of the so called Bogoliubov waves:
\begin{equation}
v_g(k)=\frac{k^2/2+c_0^2}{\sqrt{k^2/4+c_0^2}},
\end{equation}
and here $k=2\pi/L=2\sqrt{\la_{\sss\rm H}^2 - c_0^2}$
[where $L$ is computed from Eq.~\eqref{4-6}]. This yields
$v_g=2\,\la_{\sss\rm H} - c_0^2/\la_{\sss\rm H}=v_{\sss\rm H}$,
as it should. This property of the small-amplitude edge is
especially important in the theory of DSWs for non-integrable
equations (see, e.g., Refs.~\cite{El05,Kam18}).

The value of $W_4^{\sss\rm A}(\la_{\sss\rm H},\la_{\sss\rm H})$
  in Eq. \eqref{ek34} is computed through \eqref{ek17} and
  \eqref{ek28}. One gets
\begin{equation}
W_4^{\sss\rm A}(\la_{\sss\rm H},\la_{\sss\rm H}) =
\pi \, \varphi^{\sss\rm A}(\la_{\sss\rm H})
+ \pi \, \left( \la_{\sss\rm H} - \frac{c_0^2}{\la_{\sss\rm H}} \right)
\frac{d \varphi^{\sss\rm A}}{d \mu}(\la_{\sss\rm H})\; ,
\end{equation}
and
\begin{equation}
\begin{split}
\frac{{d}W_4^{\sss\rm A}(\la_{\sss\rm H},\la_{\sss\rm H})}{d \la_{\sss\rm H}}
& = \pi\left( 2 + \frac{c_0^2}{\la_{\sss\rm H}^2} \right)
\frac{d \varphi^{\sss\rm A}}{d \mu}(\la_{\sss\rm H})
\\
& + \pi\,\left( \la_{\sss\rm H} - \frac{c_0^2}{\la_{\sss\rm H}} \right)
\frac{d^2 \varphi^{\sss\rm A}}{d \mu^2}(\la_{\sss\rm H})\; ,
\end{split}
\label{ek36}
\end{equation}
where $\varphi^{\sss\rm A}$ is given by Eq.~\eqref{ek27}. Once
expression \eqref{ek36} has been used to obtain $\la_{\sss\rm H}(t)$
by solving Eq.~\eqref{ek35}, the position $x_{\sss\rm H}(t)$ of the
harmonic edge of the DSW is determined by \eqref{ek34}. The time
evolution of $x_{\sss\rm H}(t)$ is displayed in Fig.~\ref{fig9}.

The position of the point $x_{m}(t)$ where $\la_4 = c_m$
(cf. Fig.~\ref{fig8}) can be obtained from Eqs.~\eqref{ek17}. First,
for a given time $t$, one needs to find the corresponding value
$\la_3$, solution of the following equation
\begin{equation}
 t = \frac{ W_3(\la_3,c_m) -W_4(\la_3,c_m)}{v_4(\la_3,c_m) - v_3 (\la_3,c_m)}.
\end{equation}
Note that in the above we did not write the superscript A or B,
because this formula equally holds in both cases since it is to be
determined at the boundary between the two regions A and B of the DSW,
cf. Eq.~\eqref{ek21} and Fig.~\ref{fig8}. Then, $x_{m}(t)$ is
determined using any one of the Eqs.~\eqref{ek17}. The result is
shown in Fig.~\ref{fig9}, where the curve $x_{m}(t)$ represents the
position of the boundary between the two regions A and B at time $t$.

\subsection{The global picture}\label{sec:global}

We now compare the results of the Whitham approach with the numerical
solution of the NLS equation \eqref{eq:nls}
for the initial profile \eqref{rho_init}.

The DSW is described by Whitham method as explained in Secs.~\ref{sec:DSW}
and \ref{sec:GHM}. For this purpose one needs to
determine $\la_3$ and $\la_4$ as functions of $x$ and $t$ (whereas
$\la_{1}=-c_0$ and $\la_2=c_0$). This is performed as follows:
\begin{itemize}
\item First, we pick up a given $\la_4 \in [c_0,\la_{\sss\rm S}]$, where
  $\la_{\sss\rm S}$ is the value of $\la_4$ at the soliton edge, the point
  where the DSW is connected to the rarefaction wave (it has been explained
  in Secs.~\ref{sol-edge} and \ref{sec:Boundaries} how to compute it).

\item Second, at fixed $t$ and $\la_4$, we find the corresponding value
  $\la_3$ as a solution of the difference of equations~\eqref{ek17},
\begin{equation}\label{solve_r12}
\left( v_4 - v_3 \right)\cdot t = W_3(\la_3,\la_4) -W_4(\la_3,\la_4)
\, ,
\end{equation}
where $W_3$ and $W_4$ are computed from Eq.~\eqref{ek19}, with a
superscript A or B, as appropriate.
\item Last, the corresponding value of $x$ is determined by $x=W_3+v_3 t$
(or equivalently $x=W_4+v_4 t$).
\end{itemize}
This procedure gives, for each $\la_4 \in [c_0,\la_{\sss\rm S}]$ and
$t$, the value of $\la_3$ and $x$. In practice, it makes it possible
to associate with each $(x,t)$ a couple $(\la_3,\la_4)$. The results
confirm the schematic behavior depicted in Fig.~\ref{fig8}.

The knowledge of $\la_3(x,t)$ and $\la_4(x,t)$ completes our study and
enable us to determine, for each time $t> t_{\rm\sss WB}$, $\rho(x,t)$
and $u(x,t)$ as given by the Whitham approach, for all
$x\in\mathbb{R}^+$. Denoting as ${x}^*(t)$ the left boundary of the
hump (remember that we concentrate on the right part of the light
intensity profile, see Fig.~\ref{fig8}):
\begin{itemize}
\item[(i)] In the two regions $x\ge x_{\sss\rm H}(t)$ and $0\le x\le
  {x}^*(t)$, we have $u(x,t)=0$ and $\rho(x,t)=\rho_0$.

\item[(ii)] In the dispersionless region
  $[{x}^*(t),x_{\sss\rm S}(t)]$, $u(x,t)$ and $\rho(x,t)$ are computed
  from \eqref{3-3a} in terms of $\la^+$ and $\la^-$ which themselves
  are computed as explained in Sec.~\ref{DSE}. The profile in this
  region rapidly evolves to a rarefaction wave (with $\la^-=-c_0$, see
  Fig.~\ref{fig8}) of triangular shape.

\item[(iii)] Inside the DSW, for $x\in[x_{\sss\rm S}(t),x_{\sss\rm H}(t)]$,
the functions $\rho(x,t)$ and $u(x,t)$ are given by the expression
\eqref{4-2}, with $\la_1=-c_0=-\la_2$ and $\la_3$ and $\la_4$ determined as
functions of $x$ and $t$ by the procedure just explained.
\end{itemize}

\begin{figure}
\centering
\includegraphics[width=\linewidth]{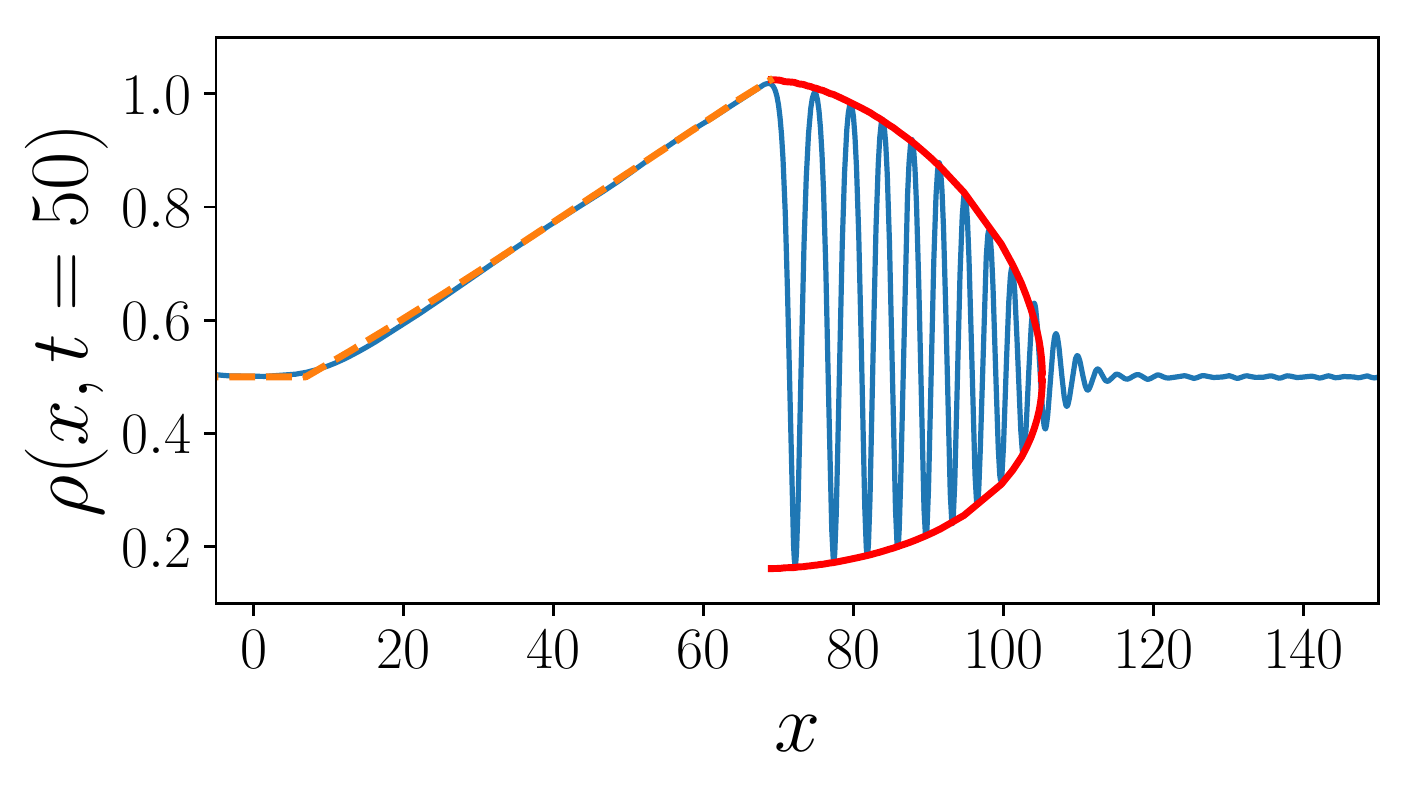} \\

\includegraphics[width=\linewidth]{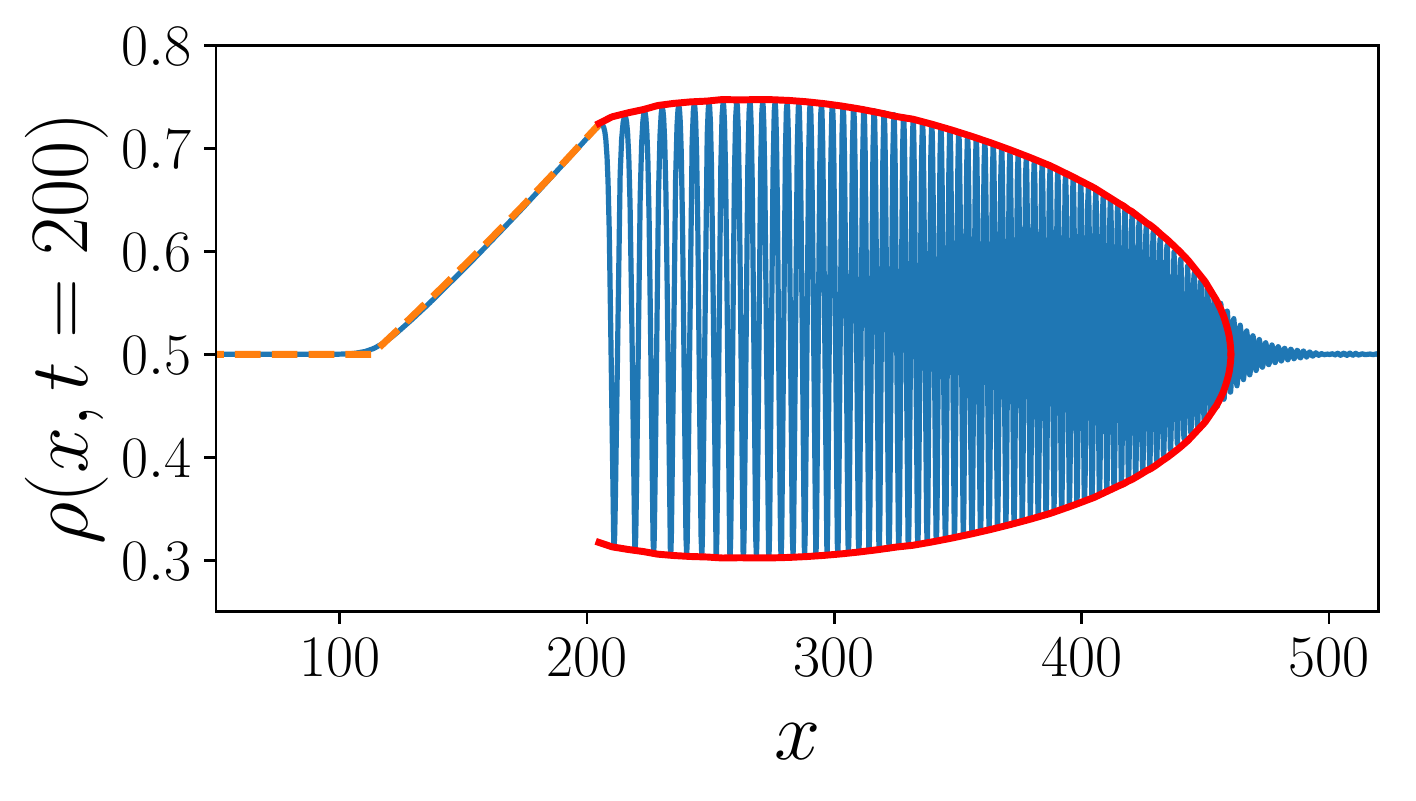}
\caption{Comparison between theory and numerical simulations for the
  density profile $\rho(x,t)$ at $t =50$ (upper plot) and $t=200$
  (lower plot). The initial profile is the same that was used in all
  the previous figures.  The blue curves are the numerical
  results. The red solid lines are the envelopes of the density
  \eqref{4-2} where the $\la_i$'s are calculated by the procedure
  described in Sec.~\ref{sec:global}. The dashed orange lines
  correspond to the the dispersionless part of the profile,
  determined using the method exposed in Sec.~\ref{DSE}.}
\label{fig11a}
\end{figure}

\begin{figure}
\centering
\includegraphics[width=\linewidth]{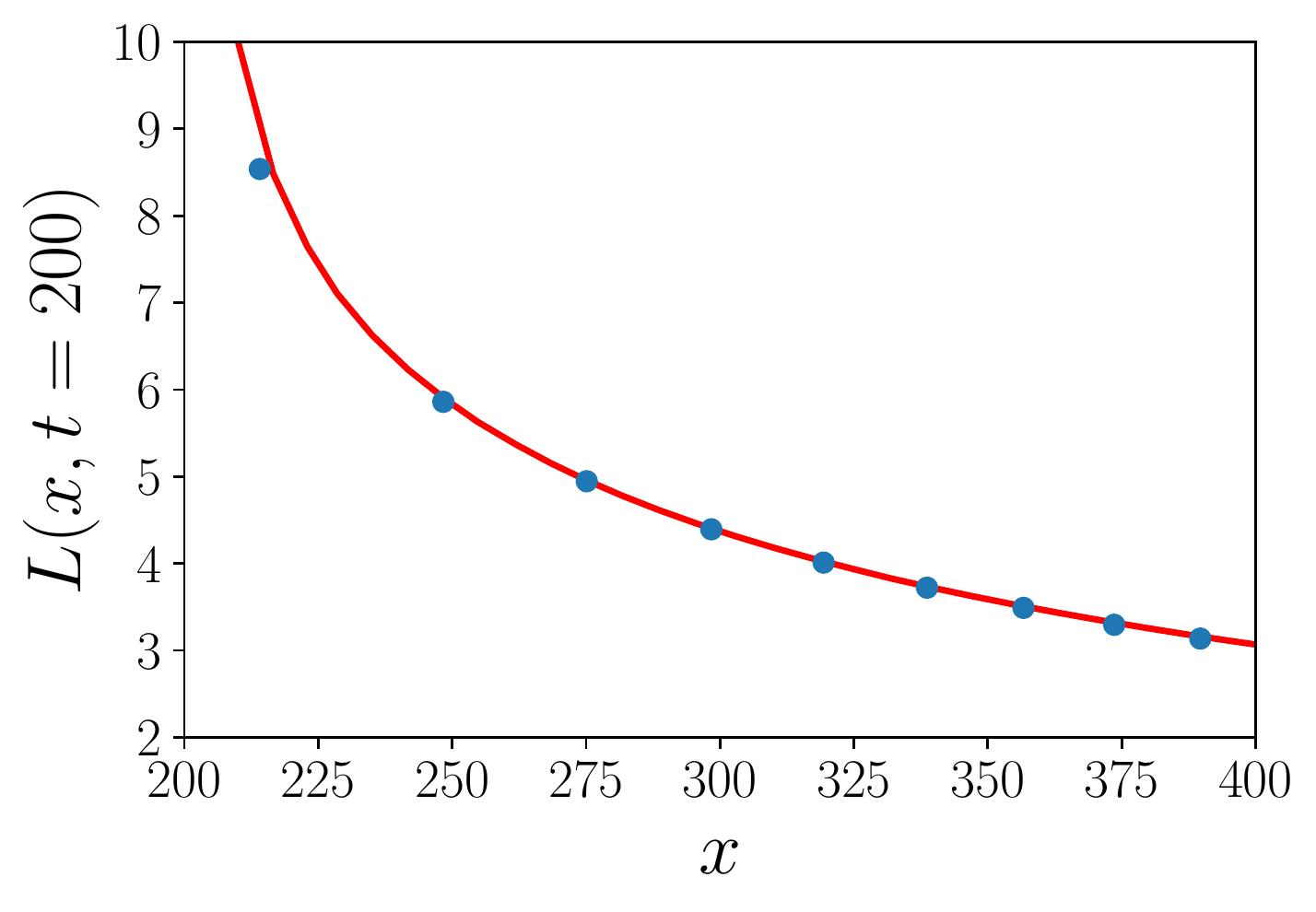}
\caption{Wavelength of the nonlinear oscillations within the DSW for
  $t=200$. The theoretical red curve is calculated from Eq.~\eqref{4-6}.
  The blue points are extracted from simulations.}
\label{fig12}
\end{figure}

The corresponding density profiles are shown in Fig.~\ref{fig11a} at
different values of time for the initial distribution \eqref{rho_init}
(with $\rho_0=0.5$, $\rho_1=1.5$ and $x_0=20$). The agreement with the
numerical simulation is excellent. The same level of accuracy is
reached for the velocity profile $u(x,t)$.

In Fig.~\eqref{fig12} we also compare the wavelength of the
nonlinear oscillations within the DSW as determined by Whitham
approach [Eq.~\eqref{4-6}]
with the results of numerical simulations, and the agreement
is again very good.

\section{Discussion and experimental considerations}\label{discu}

\begin{figure}
\centering
\includegraphics[width=\linewidth]{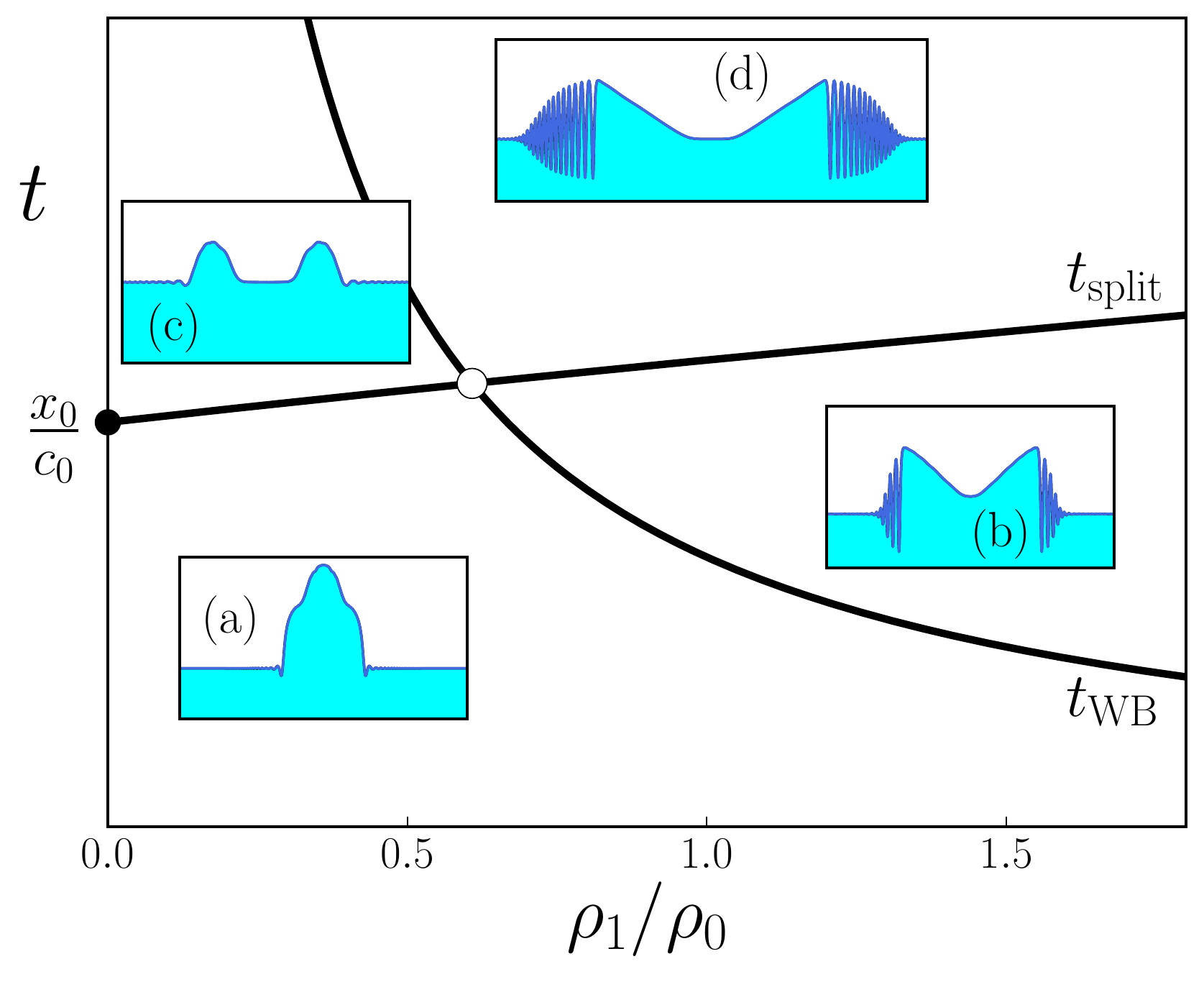}
\caption{Behavior of the light intensity profile in the plane
  $(\rho_1/\rho_0,t)$. The plane is separated in four regions by the
  curves $t=t{\rm\sss WB}$ and $t=t_{\rm split}$. These curves cross
  at the point represented by a white dot (of coordinates
  $\rho_1/\rho_0=0.60814$ and $c_0\,t/x_0=1.09623$).  Typical profiles
  are displayed in the insets (a), (b), (c) and (d) which represent
  $\rho(x,t)$ plotted as a function of $x$ for fixed $t$. }
\label{fig14}
\end{figure}

The different situations we have identified are summarized in
Fig.~\ref{fig14} which displays several typical density profiles in a
``phase space'' with coordinates $\rho_1/\rho_0$ and $t$. The curves
$t_{\rm split}(\rho_1/\rho_0)$ [as given by Eq.~\eqref{ts3}] and
$t_{\rm\sss WB}(\rho_1/\rho_0)$ [Eq. \eqref{eq:tWB_approx}] separate
this plane in four regions, labeled as (a), (b), (c) and (d) in the
figure. These two curves cross at a point represented by a white dot
whose coordinates we determined numerically as being
$\rho_1/\rho_0=0.60814$ and $c_0\,t/x_0=1.09623$. These coordinates
are universal in the sense that they have the same value for any
initial profile of inverted parabola type, such as given by
Eq.~\eqref{rho_init}, with $u(x,0)=0$. Other types of initial profile
would yield different precise arrangements of these curves in phase
space, but we expect the qualitative behavior illustrated by
Fig.~\ref{fig14} to be generic, because the different regimes depicted in
this figure correspond to physical intuition: a larger initial
hump (larger $\rho_1/\rho_0$) experiences earlier wave breaking, and
needs a longer time to be separated in two contra-propagating pulses.
Also, the evolution of a small initial pulse can initially be
described by perturbation theory and first splits in two humps which
experience wave breaking in a later stage (as illustrated in
Fig. \ref{fig1}): this is the reason why $t_{\rm split}<t_{\rm\sss WB}$
for small $\rho_1/\rho_0$. In the opposite situation where $t_{\rm\sss
  WB}<t<t_{\rm split}$, the wave breaking has already occurred while
the profile has not yet split in two separated humps. This is the
situation represented by the inset (b) and which has been considered
in Refs.~\cite{wkf-07} and \cite{Xu16}. 

\begin{figure}
\centering
\includegraphics[width=\linewidth]{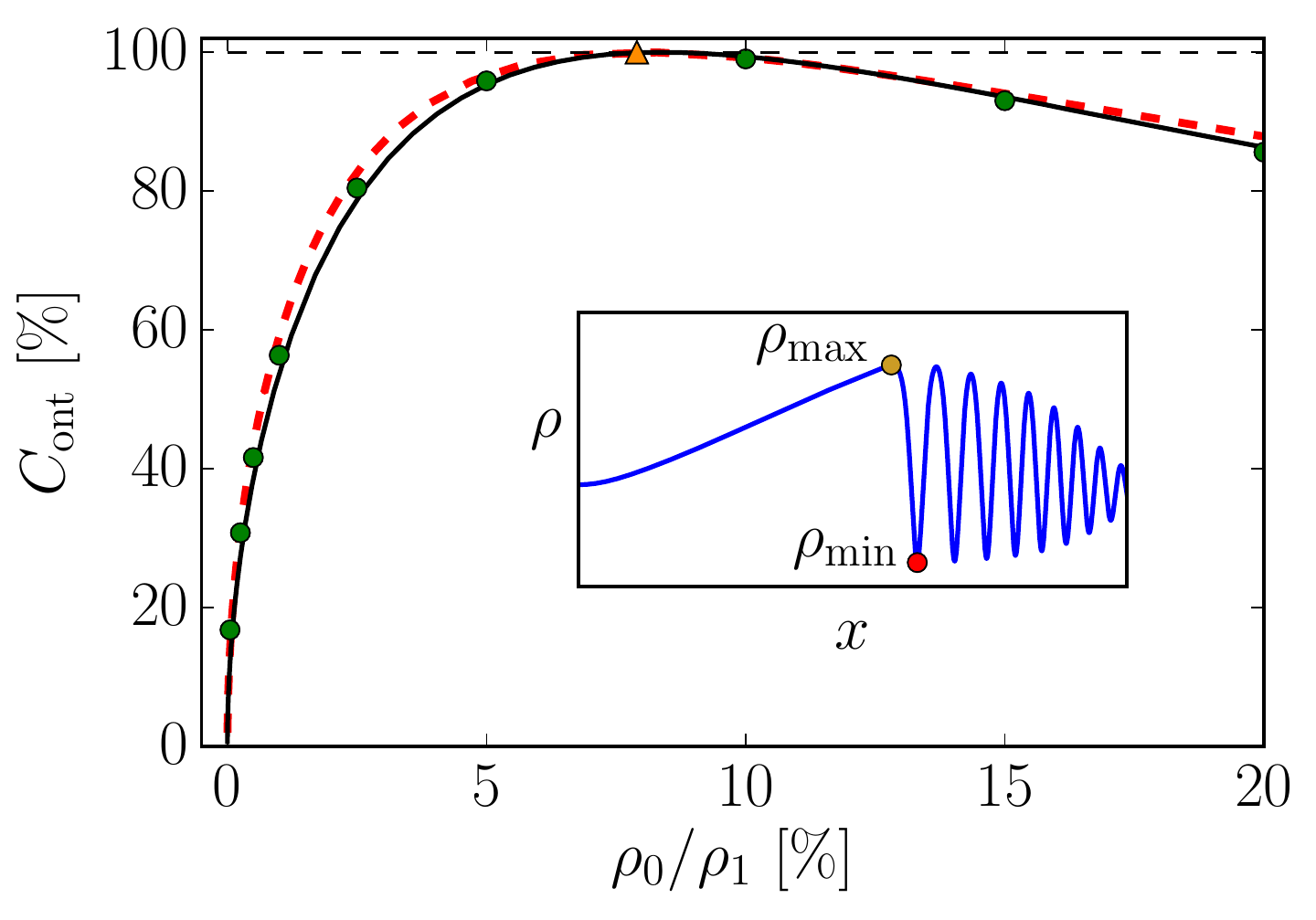}
\caption{Contrast $C_{\rm ont}$ represented as a function of
  $\rho_0/\rho_1$. We follow here the procedure of Ref. \cite{Xu16}
  and use the same dimensionless parameters: the value of $\rho_0$
  varies while $\rho_1=5.9$, $x_0=6.3$ and $t=9$ are fixed. The green
  dots correspond to the numerically determined value of the contrast,
  obtained from Eq.~\eqref{eq.cont} where $\rho_{\rm min}$ and
  $\rho_{\rm max}$ are defined as illustrated in the inset. The black
  solid line corresponds to expression \eqref{cont1}, where $\lsol$ is
  obtained from \eqref{t3-148.8a}. The red dashed line is the
  approximate result obtained from the same Eq. \eqref{cont1}, but
  evaluating $\lsol$ from Eqs. \eqref{xrrr}, \eqref{cont3} and
  \eqref{cont5}. The triangle marks the point of contrast unity.}
\label{fig15}
\end{figure}

In Ref.~\cite{Xu16}, Xu {\it et al.} studied the formation of a DSW in
a nonlinear optic fiber \cite{remf0} varying the intensity of the
background.  In particular, they quantitatively evaluated the
visibility of the oscillations near the solitonic edge of the DSW by
measuring the contrast
\begin{equation}\label{eq.cont}
C_{\rm ont} =
\frac{\rho_{\rm max}-\rho_{\rm min}}{\rho_{\rm max}+\rho_{\rm min}}\; ,
\end{equation}
where $\rho_{\rm max}$ and $\rho_{\rm min}$ are defined in the inset
of Fig.~\ref{fig15}. In Ref.~\cite{Xu16}, the contrast was studied for
a fiber of fixed length, for an initial Gaussian bump --- i.e.,
different from \eqref{rho_init} --- keeping the quantities analogous to
$\rho_1$ and $x_0$ fixed and varying $\rho_0$. The experimental
results agreed very well with numerical simulations taking into
account absorption in the fiber. Here, we do not consider exactly the
same initial profile and do not take damping into account, but we show
that our approach gives a very reasonable analytic account of the
behavior of $C_{\rm ont}$ considered as a function of $\rho_0/\rho_1$.

From Eq.~\eqref{4-2} in the limit $m\to 1$ (which is the relevant regime
near the solitonic edge of the DSW) one gets
\begin{equation}\label{cont0}
\rho_{\rm max}=\tfrac14 (\lsol+c_0)^2\;,\; \; \mbox{and}\quad
\rho_{\rm min}=\tfrac14 (\lsol-3 c_0)^2\; ,
\end{equation}
yielding
\begin{equation}\label{cont1}
C_{\rm ont}=\frac{4 c_0 (\lsol-c_0)}{(\lsol-c_0)^2+4 c_0^2}\; .
\end{equation}
The results presented in Fig.~\ref{fig15} demonstrate that, as
expected, this expression (black solid line in the figure) agrees very
well with the contrast determined from the numerical solution of
Eq.~\eqref{eq:nls} (green dots).

At this point, the computation of $C_{\rm ont}$ through \eqref{cont1}
relies on the determination of $\lsol$ by means of \eqref{t3-148.8a},
a task which requires a good grasp of the Riemann approach. However,
one can get an accurate, though approximate, analytic determination of
$C_{\rm ont}$ in a simpler way: by using the large time expression
\eqref{xrrr} for $\lsol$, together with the approximation
\begin{equation}\label{cont2}
\begin{split}
 \cal A\simeq&
- \Bigg(\int_{c_m}^{c_0}\sqrt{r-c_0}\,\frac{dw^{\sss\rm B}(r)}{d r}dr\\
  &+\int_{c_0}^{c_m}\sqrt{r-c_0}\,\frac{dw^{\sss\rm A}(r)}{d r}dr\Bigg)\\
=&\; 2 \int_0^{x_0}\sqrt{\la^+(x,0)-c_0} \; dx.
\end{split}
\end{equation}
In the above, we approximated in expression \eqref{bigA}
$W_{+}^{(3/2)}(-c_0,r)$ by $w^{\sss\rm A/B}(r)$, used the symmetry of
these functions and made the change of variable
$x=w^{\sss\rm A}(r)\Leftrightarrow r=\la^+(x,0)$, in which
$\la^+(x,0)=\sqrt\rho(x,0)$ where $\rho(x,0)$ is the initial density
profile \eqref{rho_init}. A new
change of variable yields
\begin{equation}\label{cont3}
{\cal A}\simeq 2 x_0\sqrt{c_0} \, F(\rho_0/\rho_1)\; ,
\end{equation}
where
\begin{equation}\label{cont4}
F(\alpha)=\int_0^{\pi/2}
\cos\theta\left(\sqrt{1+\frac{\cos^2\theta}{\alpha}}-1\right)^{1/2}
\!\! d\theta.
\end{equation}
A simple analytic expression of $F(\alpha)$ cannot be obtained, but
we checked that one can devise an accurate approximation by expanding
the term in parenthesis in the above integrand around $\theta=0$
up to second order in
$\theta$. This yields
\begin{equation}\label{cont5}
\begin{split}
F(\alpha)\simeq& \;
\frac{(\sqrt{\alpha+1}-\sqrt{\alpha})^{1/2}}{\alpha^{1/4}}\\
& -
\frac{\tfrac14(\pi^2/4-2)}
{\alpha^{1/4}\sqrt{1+\alpha}(\sqrt{\alpha+1}-\sqrt{\alpha})^{1/2}}\; .
\end{split}
\end{equation}
In the domain $10^{-3}\le \alpha\le 50$, $F(\alpha)$ varies over two
orders of magnitude (from 4.8 to $7.8\times10^{-2}$), and
the approximation \eqref{cont5} gives an absolute error ranging from
$5.8\times 10^{-2}$ to $1.8\times 10^{-3}$, and a relative one
ranging from 1.1 \% to 2.4 \%.

Combining Eqs.~\eqref{cont1}, \eqref{xrrr}, \eqref{cont3} and
\eqref{cont5} yields an analytic expression for the contrast
$C_{\rm ont}$. This expression is represented as a dashed red line in
Fig.~\ref{fig15}. As one can see, it compares quite well with the
value of $C_{\rm ont}$ extracted from the numerical simulations
\cite{remf}. The better agreement with the numerical result is reached
for small $\rho_0/\rho_1$; this was expected: in this regime the wave
breaking occurs rapidly, and one easily fulfils the condition
$t\gg t_{\rm\sss WB}$ where the approximation \eqref{xrrr} holds.  We
note here that the behavior of the contrast illustrated in
Fig.~\ref{fig15} is very similar to the one obtained in
Ref.~\cite{Xu16}. In both cases there is a special value of
$\rho_0/\rho_1$ for which the contrast is unity, meaning that the
quantity $\rho_{\rm min}$ cancels. From \eqref{cont0} and \eqref{xrrr}
this is obtained for $2 c_0\simeq ({\cal A}/t)^{2/3}$, i.e. - using
\eqref{cont3} -- for
\begin{equation}\label{cont6}
\frac{c_0\, t}{x_0}=\frac{1}{\sqrt{2}}\, F(\rho_0/\rho_1)\; .
\end{equation}
A numerical solution of this equation gives, for the parameters of
Fig.~\ref{fig15}, a contrast unity when $\rho_0/\rho_1=7.9$ \%, while
the exact Eq.~\eqref{cont1} predicts a maximum contrast when
$\rho_0/\rho_1=8.3$ \% instead (the exact result at
$\rho_0/\rho_1=7.9$ \% is $C_{\rm ont}=0.999$). These two values are
marked with a single triangle in Fig. \ref{fig15} because they cannot
be distinguished on the scale of the figure. This shows that the
solution of Eq.~\eqref{cont6} gives a simple way for determining the
best configuration for visualizing the fringes of the DSW; this should
be useful for future experimental studies.

Note that formula \eqref{cont1} demonstrates that the contrast depends only on
$\lsol/c_0$, and using the approximate relations
\eqref{xrrr} and \eqref{cont6} leads to the conclusion that $C_{\rm
  ont}$ can be considered as a function of the single variable
\begin{equation}
X=\frac{x_0}{t\sqrt{\rho_1}}\sqrt{\frac{\rho_1}{\rho_0}}F(\rho_0/\rho_1)\; .
\end{equation}
Hence, for a configuration different from the one considered in
Fig. \ref{fig15} but for which the combination of parameters
$t\, \sqrt{\rho_1}/x_0$ takes the same value (namely 3.47), the curve
$C_{\rm cont}(\rho_0/\rho_1)$ should superimpose with the one
displayed in Fig. \ref{fig15}. We checked that this is indeed the case
by taking $\rho_1=2$, $x_0=20$ and $t=49$, but did not plot the
corresponding contrast in Fig. \ref{fig15} for legibility.

Fig.~\ref{fig15} and the discussion of this section illustrate the
versatility of our approach which, not only gives an excellent account
of the numerical simulations at the prize of an elaborate
mathematical treatment, but also provides simple limiting expressions
--- such as Eq.~\eqref{xrrr} --- which make it possible to obtain an
analytic and quantitative description of experimentally relevant
parameters such as the contrast of the fringes of the DSW.

\section{Conclusion}\label{conclusion}

In this work we presented a detailed theoretical treatment of the
spreading of a light pulse propagating in a nonlinear medium. A
hydrodynamic approach to both the initial nondispersive spreading and
the subsequent formation of an optical dispersive shock compares
extremely well with the results of numerical simulations. Although in
reality the transition between these two regimes is gradual, it is
sharp within the Whitham approximation.  An exact expression has been
obtained for the theoretical wave breaking time which separates these
two regimes [Eq.~\eqref{eq:tWB_approx}], which may be used for
evaluating the experimental parameters necessary for observing a DSW
in a realistic setting (see Fig. \ref{fig14}). Besides, our
theoretical treatment provides valuable insight into simple features
of the shocks which are relevant to future experimental studies, such
as the coordinates of its trailing edge $\xsol$, the large-time
nondispersive intensity profile which follows it
(Sec.~\ref{sol-edge}), and the best regime for visualizing the fringes
of the DSW (Sec.~\ref{discu}).  We note also that our treatment
reveals the existence of an asymptotically conserved quantity which
had remained unnoticed until now, see Eq.~\eqref{conserved}.

A possible extension of the present work would be to consider an
initial configuration for which, at variance with the situation we
study here, the largest intensity gradient is not reached exactly at
the extremity of the initial hump. In this case, wave breaking occurs
within a simple wave (not at its boundary), and the DSW has to be
described by four position and time dependent Riemann invariants
\cite{ek-95}. In vicinity of the wave breaking moment, one of the
Riemann invariants can be considered as constant, and a generic
dispersionless solution can be represented by a cubic parabola; for
this simpler case the detailed theory was developed in
Ref.~\cite{kku-02}.  In Refs.~\cite{ekv-01,Gra07} the general
situation was considered for the Korteweg-de Vries equation.

We conclude by stressing that the present treatment focuses on
quasi-one dimensional spreading; future developments should consider
non exactly integrable systems, for instance light propagation in a
photorefractive medium, in a bi-dimensional situation with
cylindrical symmetry. Work in these directions is in progress.

\begin{acknowledgments}
  We thank T. Bienaim\'e, Q. Fontaine and Q. Glorieux for fruitful
  exchanges. We also thank an anonymous 
referee for suggestions of improvement of the manuscript. A. M. K. thanks Laboratoire de Physique Th\'eorique et
  Mod\`eles Statistiques (Universit\'e Paris-Saclay) where this work
  was started, for kind hospitality. This work was supported by the
  French ANR under Grant No. ANR-15-CE30-0017 (Haralab project).
\end{acknowledgments}


\begin{thebibliography}{00}

\bibitem{Tal65} V. I. Talanov, Radiophys. {\bf 9}, 138 (1965).

\bibitem{ASK66} S. A. Akhmanov, A. P. Sukhorukov and R. V. Khokhlov,
  Zh. Eksp. Teor. Fiz. {\bf 50}, 1537 (1966) [Sov. Phys. JETP {\bf
    23}, 1025 (1966)].

\bibitem{ASK67} S. A. Akhmanov, A. P. Sukhorukov and R. V. Khokhlov,
Usp. Fiz. Nauk {\bf 93}, 19 (1967)
[Sov. Phys. Usp. {\bf 10}, 609 (1968)] \doi{10.1070/PU1968v010n05ABEH005849}

\bibitem{Akh67} S. A. Akhmanov, D. P. Krindach, A. P. Sukhorukov, and
  R. V. Khokhlov Pis'ma Zh. Eksp. Teor. Fiz. {\bf 6}, 509 (1967) [JETP
  Lett. {\bf 6}, 38 (1967)].

\bibitem{Har97} R.G. Harrison, L. Dambly, Dejin Yu, Weiping Lu,
Opt. Commun. {\bf 139}, 69 (1997)
\doi{10.1016/S0030-4018(96)00790-0}

\bibitem{Emp87} P. Emplit, J. P. Hamaide, F. Reynaud, C. Froehly, and
  A. Barthelemy, Opt. Commun. {\bf 62}, 374 (1987)
  \doi{10.1016/0030-4018(87)90003-4}

\bibitem{Kro88} D. Kr\"okel, N. J. Halas, G. Giuliani, and D. Grischkowsky
Phys. Rev. Lett. {\bf 60}, 29 (1988) \doi{10.1103/PhysRevLett.60.29}

\bibitem{Wei88} A. M. Weiner, J. P. Heritage, R. J. Hawkins,
  R. N. Thurston, E. M. Kirschner, D. E. Leaird, and W. J. Tomlinson,
  Phys. Rev. Lett. {\bf 61}, 2445 (1988) \doi{10.1103/PhysRevLett.61.2445}

\bibitem{Are91} F. T. Arecchi, G. Giacomelli, P. L. Ramazza, and S. Residori
Phys. Rev. Lett. {\bf 67}, 3749 (1991) \doi{10.1103/PhysRevLett.67.3749}

\bibitem{Vau96} M. Vaupel, K. Staliunas, and C. O. Weiss
Phys. Rev. A {\bf 54}, 880 (1996) \doi{10.1103/PhysRevA.54.880}

\bibitem{Voc18} D. Vocke, K. Wilson, F. Marino, I. Carusotto,
  E. M. Wright, T. Roger, B. P. Anderson, P. \"Ohberg, D. Faccio,
  Phys. Rev. A {\bf 94}, 013849 (2016) \doi{10.1103/PhysRevA.94.013849}.

\bibitem{Rot89} J. E. Rothenberg, and D. Grischkowsky,
Phys. Rev. Lett. {\bf 62}, 531 (1989) \doi{10.1103/PhysRevLett.62.531}

\bibitem{Gho07} N. Ghofraniha, C. Conti, G. Ruocco, and S. Trillo,
Phys. Rev. Lett. {\bf 99}, 043903 (2007) \doi{10.1103/PhysRevLett.99.043903}

\bibitem{Cou04} G. Couton, H. Maillotte, and M. Chauvet, J. Opt. B: Quantum
Semiclassical Opt. {\bf 6}, S223 (2004), \doi{10.1088/1464-4266/6/5/009}.

\bibitem{wkf-07} W. Wan, S. Jia, and J. W. Fleischer, Nature
Phys. {\bf 3}, 46 (2007), \doi{10.1038/nphys486}.

\bibitem{Bar07} C. Barsi, W. Wan, C. Sun, and J. W. Fleischer, Opt.
Lett. {\bf 32}, 2930 (2007), \doi{10.1364/OL.32.002930}


\bibitem{Fat14} J. Fatome, C. Finot, G. Millot, A. Armaroli, and
  S. Trillo, Phys. Rev. X {\bf 4}, 021022 (2014)
  \doi{10.1103/PhysRevX.4.021022}.

\bibitem{Xu16} G. Xu, A. Mussot, A. Kudlinski, S. Trillo, F. Copie,
  and M. Conforti, Opt. Lett. {\bf 41}, 2656 (2016),
  \doi{10.1364/OL.41.002656}

\bibitem{Xu17} G. Xu, M. Conforti, A. Kudlinski, A.
  Mussot, S. Trillo, Phys. Rev. Lett. {\bf 118}, 254101 (2017)
  \doi{10.1103/PhysRevLett.118.254101}

\bibitem{Kar15} M. Karpov, T. Congy, Y. Sivan, V. Fleurov, N. Pavloff
  and S. Bar-Ad, Optica {\bf 2}, 1053 (2015) \doi{10.1364/OPTICA.2.001053}

\bibitem{Ela12} M. Elazar, V. Fleurov, S. Bar-Ad, Phys. Rev. A {\bf 86},
063821 (2012)
\doi{10.1103/PhysRevA.86.063821}

\bibitem{Voc17} D. Vocke, C. Maitland, A. Prain, F. Biancalana,
  F. Marino, E. M. Wright, D. Faccio, Optica {\bf 5}, 1099 (2018)
  \doi{10.1364/OPTICA.5.001099}

\bibitem{Dro18} J. Drori, Y. Rosenberg, D. Bermudez, Y. Silberberg,
  and U. Leonhardt, Phys. Rev. Lett. {\bf 122}, 010404 (2019),
  \doi{10.1103/PhysRevLett.122.010404}

\bibitem{Voc15} D. Vocke, T. Roger, F. Marino, E. M. Wright,
  I. Carusotto, M. Clerici, and D. Faccio, Optica {\bf 2}, 484 (2015)
  \doi{10.1364/OPTICA.2.000484}

\bibitem{Fon18} Q. Fontaine, T. Bienaim\'e, S. Pigeon, E. Giacobino, A.
Bramati, and Q. Glorieux, Phys. Rev. Lett. {\bf 121}, 183604 (2018)
\doi{10.1103/PhysRevLett.121.183604}.

\bibitem{Mic18} C. Michel, O. Boughdad, M. Albert, P.-\'E. Larr\'e, and
  M. Bellec, Nat. Commun. {\bf 9}, 2108 (2018)
  \doi{10.1038/s41467-018-04534-9}

\bibitem{kgk-04} A. M. Kamchatnov, A. Gammal, and R. A. Kraenkel,
  Phys. Rev. A {\bf 69}, 063605 (2004) \doi{10.1103/PhysRevA.69.063605}.

\bibitem{El07} G. A. El, A. Gammal, E. G. Khamis, R. A. Kraenkel, and
  A. M. Kamchatnov, Phys. Rev. A {\bf 76}, 053813 (2007)
  \doi{10.1103/PhysRevA.76.053813}

\bibitem{Conf12} M. Conforti and S. Trillo in {\it Rogue and Shock
    Waves in Nonlinear Dispersive Media}, M. Onorato, S. Residori and
  F. Baronio eds., Lecture Notes in Physics Volume 926, (Springer
  International Publishing, Switzerland 2016).

\bibitem{whitham-74} G.~B.~Whitham, {\it Linear and Nonlinear Waves}
  (Wiley Interscience, New York, 1974).

\bibitem{Iso18} M. Isoard, A. M. Kamchatnov, and N. Pavloff,
  Phys. Rev. E {\bf 99}, 012210 (2019) \doi{10.1103/PhysRevE.99.012210}

\bibitem{Lud52} G. S. S. Ludford, Proc. Camb. Phil. Soc. {\bf 48}, 499
  (1952) \doi{10.1017/S0305004100027900}

\bibitem{For2009} M. G. Forest, C.-J. Rosenberg, and O. C. Wright III,
Nonlinearity {\bf 22}, 2287 (2009), \doi{10.1088/0951-7715/22/9/012}


\bibitem{gp-73} A. V. Gurevich and L. P. Pitaevskii,
  Zh. Eksp. Teor. Fiz. {\bf 65}, 590 (1973) [Sov. Phys. JETP {\bf 38},
  291 (1974)].

\bibitem{eh-16} G. A. El and M. A. Hoefer, Physica D {\bf 333}, 11
  (2016), \doi{10.1016/j.physd.2016.04.006}

\bibitem{gkm-89} A. V. Gurevich, A. L. Krylov, and N. G. Mazur,
  Zh. Eksp. Teor. Fiz. {\bf 95}, 1674 (1989) [Sov. Phys. JETP {\bf
    68}, 966 (1989)].

\bibitem{Gur91} A. V. Gurevich, A. L. Krylov and G. A. El, Pis'ma
  Zh. Eksp. Teor. Fiz. {\bf 54}, 104 (1991) [JETP Lett. \textbf{54},
  102 (1991)].

\bibitem{Gur92} A. V. Gurevich, A. L. Krylov and G. A. El,
  Zh. Eksp. Teor. Fiz. {\bf 101}, 1797 (1992) [Sov. Phys. JETP
  \textbf{74}, 957 (1992)].

\bibitem{Kry92} A. L. Krylov, V. V. Khodorovskii and G. A. El, Pis'ma
  Zh. Eksp. Teor. Fiz. {\bf 56}, 325 (1992) [JETP Lett. \textbf{56},
  323 (1992)].

\bibitem{ek-93} G. A. El and V. V. Khodorovsky, Phys. Lett. A {\bf
    182}, 49 (1993), \doi{10.1016/0375-9601(93)90051-Z}.

\bibitem{El2009} G. A. El, A. M. Kamchatnov, V. V. Khodorovskii, E. S.
  Annibale, and A. Gammal, Phys. Rev. E {\bf 80}, 046317 (2009),
  \doi{10.1103/PhysRevE.80.046317}.

\bibitem{LL8} L. D. Landau and E. M. Lifshitz, {\it Electrodynamics of
    Continuous Media}, Course of Theoretical Physics vol. 8 (Elsevier
  Butterworth-Heinemann, Oxford, 2006).

\bibitem{Tsa91} S. P. Tsarev, Math. USSR Izv. {\bf 37}, 397 (1991),
\doi{10.1070/IM1991v037n02ABEH002069}.

\bibitem{Som64} A. Sommerfeld, {\it Partial Differential Equations in Physics},
(Lectures on Theoretical Physics volume VI) (Academic Press, New York, 1964).

\bibitem{AS} M. Abramowitz and I. A. Stegun, {\it Handbook of
    mathematical functions''}, (Dover Publications, New York, 1970).

\bibitem{kamch-2000} A. M. Kamchatnov, {\it Nonlinear Periodic Waves
    and Their Modulations---An Introductory Course}, (World
  Scientific, Singapore, 2000).

\bibitem{LL-6} L. D. Landau and E. M. Lifshitz, {\it Fluid Mechanics},
  (Pergamon, Oxford, 1987).

\bibitem{foot1} It should not be a surprise that, for determining the
  wave breaking time, it is legitimate to replace
  $W^{(3)}_+(-c_0,\lambda^+)$ by $w^{\rm\sss A}(\lambda^+)$ in
  Eq. \eqref{tWB-est0}. Indeed, the wave breaking phenomenon occurs
  when several characteristics issued from the right region of the
  edge of the bump cross (in region III for the case considered). For
  all these characteristics, one intuitively expects that one can
  neglect the weak dependence of $\lambda^-$ on $x$; this is indeed
  the case: one can rigorously show that $W_+^{(3)}(-c_0,\lambda^+)$ and
  $w^{\rm\sss A}(\lambda^+)$ coincide in the limit $\lambda^+=c_0$,
  which is the relevant one for evaluating \eqref{tWB-est0}.

\bibitem{fl86} M. G. Forest and J. E. Lee, Geometry and modulation
  theory for periodic nonlinear Schr\"odinger equation, in {\it
    Oscillation Theory, Computation, and Methods of Compensated
    Compactness,} Eds. C. Dafermos {\it et al.}, IMA Volumes on
  Mathematics and its Applications {\bf 2}, p. 35 (Springer, N.Y., 1986).
  \doi{10.1007/978-1-4613-8689-6_3}.

\bibitem{pavlov87} M. V. Pavlov, Teor. Mat. Fiz. {\bf 71}, 351 (1987) [Theoret.
Math. Phys. {\bf 71}, 584 (1987)] \doi{10.1007/BF01017090}.

\bibitem{Moi63} R. Z. Sagdeev, Cooperative Phenomena and Shock Waves
  in Collisionless Plasmas, in {\it Reviews of Plasma Physics,}
  Ed. M. A. Leontovich, Vol.~4, p.~23, (Consultants Bureau, New York,
  1966).
  
\bibitem{ek-95} G. A. El and A. L. Krylov, Phys. Lett. {\bf 203}, 77 (1995)
\doi{10.1016/0375-9601(95)00379-H}.

\bibitem{wright} O. C. Wright, Commun. Pure Appl. Math. \textbf{46},
  423 (1993) \doi{10.1002/cpa.3160460306}

\bibitem{tian} F. R. Tian, Commun. Pure Appl. Math. \textbf{46}, 1093 (1993)
\doi{10.1002/cpa.3160460802}.

\bibitem{Eis18} L. P. Eisenhart, Ann. Math. {\bf 120}, 262 (1918).

\bibitem{Ark05} G. Arfken and H. J. Weber, {\it Mathematical Methods for
  Physicists} (Academic Press, Orlando, 2005).

\bibitem{El05} G. A. El, Chaos {\bf 15}, 037103 (2005),
  \doi{10.1063/1.1947120}; {\it ibid.} {\bf 16}, 029901 (2006),
  \doi{10.1063/1.2186766}.

\bibitem{Kam18} A. M. Kamchatnov, Phys. Rev. E {\bf 99}, 012203 (2019)
\doi{10.1103/PhysRevE.99.012203}

\bibitem{remf0} In
this case the role of variable $x$ in Eq. \eqref{eq:nls} is played by
time, but the phenomenology is very similar to the one we
describe in the present work, see, e.g., Y. S. Kivshar and G. P. Agrawal,
 {\it Optical Solitons} (Academic Press, San Diego, 2003).

\bibitem{remf} Computing the contrast using expression
  \eqref{cont4} instead of the approximation \eqref{cont5} yields a
  result which is barely distinguishable from the dashed line in
  Fig.~\ref{fig15}.
  
\bibitem{kku-02} A. M. Kamchatnov, R. A. Kraenkel and B. A. Umarov,
Phys. Rev. E {\bf 66}, 036609 (2002) \doi{10.1103/PhysRevE.66.036609}.

\bibitem{ekv-01} G. A. El, A. L. Krylov, and S. Venakides, Commun. Pure
  Appl. Math. {\bf 54}, 1243 (2001) \doi{10.1002/cpa.10002}.

\bibitem{Gra07} T. Grava and C. Klein, Comm. Pure Appl. Math. {\bf
    60}, 1623 (2007) \doi{10.1002/cpa.20183}.

\end{thebibliography}
\end{document}